\documentclass[12pt]{article}
\usepackage[bottom]{footmisc}
\usepackage[utf8]{inputenc}
\usepackage{authblk}
\usepackage{epsfig}
\usepackage{amssymb}
\usepackage{amsfonts}
\usepackage[mathscr]{euscript}
\usepackage{amsbsy}
\usepackage[all]{xy}
\usepackage{amsmath}
\usepackage{enumerate}
\usepackage{float}

 \usepackage{enumerate}
 \usepackage{hyperref}

\usepackage{amssymb,amscd}
\usepackage{amsmath,amsthm}
\usepackage{xspace}
\usepackage{url}
\newcommand{\del}{\partial}

\usepackage{epsfig}
\usepackage{amssymb}
\usepackage{amsfonts}
\usepackage{amsbsy}
\usepackage[all]{xy}
\usepackage{amsmath}
\usepackage{enumerate}

\usepackage{amssymb,amscd}
\usepackage{mathrsfs}
\usepackage{amsmath,amsthm}
\usepackage{xspace}
\usepackage{url}

\numberwithin{equation}{section}

\newcommand{\bea}{\begin{eqnarray}\displaystyle}
\newcommand{\eea}{\end{eqnarray}}

\newcommand{\cwarrow}{\text{\Large$\curvearrowright$}}

\newcommand\fq{\mathfrak{q}}

\newcommand\bC{\mathbb{C}}
\newcommand\bZ{\mathbb{Z}}

\newcommand\cB{\mathcal{B}}
\newcommand\cE{\mathcal{E}}

\newcommand\cH{\mathcal{H}}
\title{Categorical Pentagon Relations and Koszul Duality}
\author[a]{Davide Gaiotto\footnote{{\tt dgaiotto@perimeterinstitute.ca}}}
\author[b]{Ahsan Khan\footnote{{\tt khan@ias.edu}}}
\affil[a]{Perimeter Institute for Theoretical Physics \protect \\ 31 Caroline Street North, Waterloo ON N2L 2Y5 \protect  \\ $\,\,\,$ }
\affil[b]{School of Natural Sciences \protect \\ Institute for Advanced Study \protect \\ Einstein Drive, Princeton NJ 08540 }
\date{\today}

\begin{document}

\maketitle 

\begin{abstract} The Kontsevich-Soibelman wall-crossing formula is known to control the jumping behavior of BPS state counting indices in four-dimensional theories with $\mathcal{N}=2$ supersymmetry. The formula can take two equivalent forms: a ``fermionic'' form with nice positivity properties and a ``bosonic'' form with a clear physical interpretation. In an important class of examples, the fermionic form of the formula has a mathematical categorification involving PBW bases for a Cohomological Hall Algebra. The bosonic form lacks an analogous categorification. We construct an equivalence of chain complexes which categorifies the simplest example of the bosonic wall-crossing formula: the bosonic pentagon identity for the quantum dilogarithm. The chain complexes can be promoted to differential graded algebras which we relate to the PBW bases of the relevant CoHA by a certain quadratic duality. The equivalence of complexes then follows from the relation between quadratic duality and Koszul duality. We argue that this is a special case of a general phenomenon: the bosonic wall-crossing formulae are categorified to equivalences of $A_\infty$ algebras which are quadratic dual to PBW presentations of algebras which underlie the fermionic wall-crossing formulae. We give a partial interpretation of our differential graded algebras in terms of a holomorphic-topological version of BPS webs. \end{abstract}

\pagebreak

\tableofcontents

\pagebreak
 
\section{Introduction} 
 
\paragraph{} A given four-dimensional quantum field theory with $\mathcal{N}=2$ supersymmetry will typically have a moduli space $\cB$ of ``Coulomb branch'' vacua, so called because the infrared effective description consists of massless Abelian gauge fields together with massive dyonic matter \cite{Seiberg:1994rs}. 

\paragraph{} The Kontsevich-Soibelman wall-crossing formula controls how the spectrum of BPS particles jumps across walls of marginal stability in the space of vacua \cite{Kontsevich:2008fj,Gaiotto:2010be}. It can be compactly formulated with the help of an auxiliary ``quantum torus algebra." Letting $\Gamma$ be the charge lattice of the theory, with the anti-symmetric integral pairing $\langle \cdot, \cdot \rangle$, the quantum torus algebra is generated by a collection of formal variables $\{ \{X_{\gamma} \}_{\gamma \in \Gamma}, \fq \}$ subject to the relation\footnote{Several formulae in the draft would take a more conventional form in terms of a variable $q = \fq^2$, but we will stick to $\fq$ to avoid unnecessary square roots.}
\begin{equation}
	X_\gamma X_{\gamma'} = \fq^{\langle \gamma, \gamma'\rangle} X_{\gamma + \gamma'}.
\end{equation}
The KS formula also employs the linear ``central charge'' map $Z: \Gamma \to \bC$, such that $|Z(\gamma)|$ is the mass of a BPS particle of charge $\gamma$. In the following, we will use the notation $Z_\gamma := Z(\gamma)$. The central charge varies holomorphically as we move along $\cB$.\footnote{Recall, though, that the lattice $\Gamma$ is actually fibered non-trivially on $\cB$.}
\footnote{The notion of BPS spectrum makes sense if central charges are generic enough, namely if $\arg Z_\gamma \neq \arg Z_{\gamma'}$ whenever $\gamma$ and $\gamma'$ are not multiples of each other. This condition guarantees that the single-particle BPS spectrum is separated from the continuum of two-particle BPS states. Wall-crossing can occur at co-dimension one loci in $\cB$ where this condition fails. }

\paragraph{} The spectrum of BPS particles of charge $\gamma$ is encoded in certain elements ${\cal K}_\gamma \in \bZ[[X_\gamma,\fq]]$ which are multiplied in an order which depends on the phase of the central charge \cite{Kontsevich:2008fj}: 
\begin{equation}\label{eq:qsg}
	{\cal S}_\vartheta = \prod^{\cwarrow}_{\vartheta\leq \arg Z_\gamma<\vartheta+\pi} {\cal K}_\gamma.
\end{equation}
This ``phase-ordered" product ${\cal S}_{\vartheta}$ is known as the quantum spectrum generator. The notion of a phase-ordered product makes sense because factors with the same $\arg Z_\gamma$ commute. ${\cal S}_{\vartheta}$ itself makes sense as a formal power series in $\fq$ and in $X_\gamma$ where $\gamma$ is such that $\arg Z_\gamma \in [\vartheta, \vartheta+ \pi)$. 
 
\paragraph{} The KS wall-crossing formula states that ${\cal S}_\vartheta$ only changes when the phase of the central charge for a BPS particle leaves the interval $[\vartheta,\vartheta+ \pi)$, irrespective of how the relative orders of the phases of BPS particles vary while remaining within the interval. The statement also holds for products associated to any phase interval of width smaller than $\pi$.\footnote{We do not extend the product to a broader interval because we do not wish to multiply power series in $X_\gamma$ and $X_{-\gamma} = X_\gamma^{-1}$. } \paragraph{} For more details on the KS wall-crossing formula and related topics we refer the reader to the review articles \cite{KontsevichReview, MooreLectures, CecottiLectures}.

\paragraph{} There are two natural categorifications of the wall-crossing formula:
\begin{enumerate}
	\item If the BPS spectrum can be captured by representations of an auxiliary quiver $Q$ \footnote{See \cite{Alim:2011kw, Cecotti:2012se} and references therein.}, one may define a ``Cohomological Hall Algebra'' $\mathcal{H}(Q)$, such that the character (also known as the motivic Donaldson-Thomas series) of $\mathcal{H}(Q)$ coincides with ${\cal S}_\vartheta^{-1}$ for some $\vartheta$ and the factorization \eqref{eq:qsg} counts the elements of certain canonical PBW bases for $\mathcal{H}(Q)$ \cite{Kontsevich:2010px}.\footnote{We use the term PBW basis here in a somewhat loose sense. As explained in the main text, we refer to a collection of sub-algebras $A_\gamma$ of $\mathcal{H}(Q)$ such that products of elements $a_\gamma \in A_\gamma$ ordered by $\arg Z_\gamma$ form a linear basis for $\mathcal{H}(Q)$. } The wall-crossing invariance of $\mathcal{H}(Q)$ thus provides a mathematical categorification of the wall-crossing formula. This line of reasoning was carried out in \cite{Davison:2016bjk}. Interestingly, ${\cal K}_\gamma^{-1}$ is always a power series with positive coefficients only\footnote{This statement and its categorification is an aspect of the ``no exotics'' conjecture for 4d ${\cal N}=2$ SQFTs \cite{Gaiotto:2010be}.}, compatible with ${\cal S}_\vartheta^{-1}$ being the character of a $\Gamma \oplus \bZ$ - graded vector space\footnote{The additional $\mathbb{Z}$ factor here denotes the ``spin" grading. More precisely, the COHA is graded by the Heisenberg group $\text{Heis}(\Gamma, \langle \cdot, \cdot \rangle)$ which is a central extension of $\Gamma$ by $\mathbb{Z}$.}.  We are thus lead to conjecture that for every choice of vacuum $u \in \mathcal{B}$ and of $\vartheta$ there should be a ``BPS algebra'' $A_\vartheta$ with character ${\cal S}_\vartheta^{-1}$ and a PBW basis 
compatible with the factorization of ${\cal S}_\vartheta^{-1}$. 
	\item The ``Schur'' half-index counting supersymmetric interface local operators at certain half-BPS ``RG interfaces''\footnote{To be precise, these are local operators supported at the endpoint of half-BPS line defects in the free Abelian gauge theory at the IR side of the RG interface.} of the theory can be expressed as 
\begin{equation}
	I\!\!I(\fq) = (\fq^2;\fq^2)_\infty^{\mathrm{rk}}{\cal S}_\vartheta \, ,
\end{equation} 
where $\mathrm{rk}$ is the rank of the IR Abelian gauge group \cite{Cordova:2016uwk}. The relevant supersymmetric local operators can be expressed as local operators in a Holomorphic-Topological twist of the system \cite{Kapustin:2006hi,Aganagic:2017tvx,Costello:2020ndc}. If we could express the space of interface local operators in terms of the BPS particles of the theory, we would obtain a physical categorification of the wall-crossing formula. 
\end{enumerate}

\paragraph{} A similar phenomenon occurs in the simpler case of {\it gapped} two-dimensional $\mathcal{N}=(2,2)$ supersymmetric theories, where the wall-crossing behavior of the corresponding BPS state-counting indices is controlled by the Cecotti-Vafa formula \cite{Cecotti:1992rm}. 
There the $\gamma$ charges are replaced by soliton sectors and the ${\cal K}_\gamma$-factors are replaced by triangular matrices. 
The analogues of ${\cal S}_\vartheta$ and ${\cal S}_\vartheta^{-1}$ can be interpreted as characters for 
$A_\infty$ algebras of boundary local operators for left and right ``thimble'' boundary conditions respectively. The boundary conditions given by the direct sum of left/right thimbles are the 2d analogues of RG interfaces. These $A_\infty$ algebras are built combinatorially as ``web algebras'' \cite{Gaiotto:2015aoa}, which encode the topological twist of the system in terms of the infrared effective description involving gapped vacua, massive BPS solitons and interaction vertices of these BPS solitons. Such two-dimensional developments and their relation to categorified wall-crossing formulas are reviewed and discussed further in \cite{Khan:2020hir, Khan:2021hve}.

\paragraph{} The original goal of this work was to proceed in analogy to the two-dimensional example and build the holomorphic-topological factorization algebra of protected local operators at an RG interface in terms of some four-dimensional version of a web algebra which would manifestly demonstrate wall-crossing invariance. This goal is complicated by the fact that the 4d theory is not gapped in the Coulomb vacua. We hope that the construction in this paper captures the part of the web calculation involving the BPS particles only, without the contribution of the photons. 
Prosaically, in the examples we study we categorify  ${\cal S}_\vartheta$ instead of the whole $I\!\!I(\fq)$.  

\paragraph{} Koszul duality maps an abstract (augmented) algebra such as $\mathcal{H}(Q)$, or more generally $A_\vartheta$, with character $\mathcal{S}_{\vartheta}^{-1}$ to a relatively ``large'' $A_\infty$ algebra $A_\vartheta^!$ with character ${\cal S}_\vartheta$. If $A_\vartheta$ has extra structure, such as the PBW basis $b$ we would associate to a factorization of ${\cal S}^{-1}_\vartheta$, one can write 
an equivalent but considerably smaller ``quadratic dual'' $A_\infty$ algebra $A_\vartheta^![b]$ with a PBW basis $b^!$ which categorifies the factorization of ${\cal S}_\vartheta$. We conjecture, and demonstrate in our explicit example, that $A_\vartheta^![b]$ would also emerge as a natural ingredient of a four-dimensional version of the web algebra construction.

\subsection{The Categorified Pentagon Relation} 

\paragraph{} The simplest example of wall-crossing in four-dimensional $\mathcal{N}=2$ theories happens in the Argyres-Douglas theory of type $(A_1, A_2)$\footnote{Here we follow the $(G,G')$-notation introduced in \cite{CecottiLectures, Cecotti:2010fi}.}\cite{Argyres:2007cn,Gaiotto:2009hg}. The theory has a rank one Coulomb branch with a rank two lattice \bea \Gamma = \mathbb{Z} \langle \gamma_1 \rangle \oplus \mathbb{Z} \langle \gamma_2 \rangle \eea of electromagnetic charges, where $\gamma_1$ denotes an electric charge and $\gamma_2$ denotes a magnetic charge. The pairing on this lattice is the usual electromagnetic pairing \bea \langle \gamma_1, \gamma_2 \rangle = 1.\eea  Here there are two chambers in parameter space separated by a wall of marginal stability. In the ``strong coupling'' chamber the BPS spectrum consists of a single electric hypermultiplet of charge $\gamma_1$ and a single magnetic hypermultiplet of charge $\gamma_2$. In the ``weak coupling'' chamber the BPS spectrum consists of hypermultiplets of charges $\gamma_1, \gamma_1 + \gamma_2$ and $\gamma_2$. An additional dyon of charge $\gamma_1 + \gamma_2$ is born when crossing the wall. We will refer to these as the ``left'' and ``right'' chambers respectively. 

\paragraph{}The corresponding wall-crossing formula is given as follows. Introduce formal variables $X_{\gamma_1}, X_{\gamma_2}$ and a spin fugacity $\fq$ such that the quantum torus relations \bea X_{\gamma_1} X_{\gamma_2} = \fq X_{\gamma_1+\gamma_2} = \fq^2 X_{\gamma_2} X_{\gamma_1}\eea are obeyed. Also introduce the quantum dilogarithm 
\bea \Phi(X_{\gamma}) = \prod_{n=0}^{\infty}(1+\fq^{2n+1} X_{\gamma}) = (-\fq X_{\gamma} ;\fq^2)_{\infty}. \eea 
The wall-crossing identity for the Argyres-Douglas theory of type $(A_1, A_2)$ is \bea \label{pentagon}\Phi(X_{\gamma_2}) \Phi(X_{\gamma_1}) = \Phi(X_{\gamma_1}) \Phi(X_{\gamma_1+\gamma_2}) \Phi(X_{\gamma_2}).\eea This is known as the pentagon identity for the quantum dilogarithm \cite{Faddeev:1993rs}. Since the quantum dilogarithm $\Phi(X_{\gamma})$ coincides with the character of a fermionic Fock space, we refer to \eqref{pentagon} as the fermionic pentagon identity.  

\paragraph{} A beautiful aspect of this formula is that it holds without any cancellations: all factors are power series with positive 
coefficients. As we will discuss in Section \ref{sec:fermion}, an algebraic explanation of this fact is that the two sides count basis elements for two ``PBW bases'' for the same algebra $\mathcal{H}(A_2)$, the Cohomological Hall Algebra associated to the two-node quiver $A_2$ \cite{Kontsevich:2010px}. 

\paragraph{} The physically natural wall-crossing identity is obtained by inverting both sides of \eqref{pentagon}:
\bea \label{ipentagon} \Phi(X_{\gamma_1})^{-1} \Phi(X_{\gamma_2})^{-1}= \Phi(X_{\gamma_2})^{-1} \Phi(X_{\gamma_1+\gamma_2})^{-1} \Phi(X_{\gamma_1})^{-1} .\eea Since the inverse dilogarithm \bea \Phi(X_{\gamma})^{-1} = \prod_{n=0}^{\infty}(1+\fq^{2n+1}X_{\gamma})^{-1}\eea coincides with the character of a bosonic Fock space, we refer to \eqref{ipentagon} as the bosonic pentagon identity. 
The bosonic pentagon identity relies on heavy cancellations on the right hand side between positive and negative terms. Correspondingly, we will categorify it as a quasi-isomorphism of two chain complexes. Ultimately, we will recognize these complexes as being quasi-isomorphic formulations of a differential-graded algebra $\mathcal{H}(A_2)^!$, Koszul dual to 
$\mathcal{H}(A_2)$. 

\subsection{Outline} In Section \ref{sec:boson} we discuss and test a tentative categorification of the bosonic pentagon identity. In Section \ref{sec:fermion} we review the categorification of the fermionic pentagon identity based on COHAs. In Section \ref{sec:koszul} we review the aspects of Koszul duality relevant to us. In Section \ref{sec:dualize} we compute the Koszul dual of the COHA in two different presentations and match it with the categorification of the bosonic identity. In Section \ref{sec:web} we then interpret this differential graded algebra in a web-like language. Appendix \ref{httwist} briefly reviews the Holomorphic Topological twist of the theory of a single hypermultiplet.

\section*{Acknowledgements} We thank Kevin Costello, Tudor Dimofte and Greg Moore for useful discussions, and Dylan Allegretti for pointing out a relevant reference. AK thanks Greg Moore for piquing his interest in the problem addressed in this paper and for sharing unpublished notes (written jointly with T.~Dimofte and the first-named author) on earlier attempts. AK thanks the Perimeter Institute for Theoretical Physics for hospitality. AK is supported by the Institute for Advanced Study and the National Science Foundation under Grant No. PHY-2207584. This research was supported in part
by a grant from the Krembil Foundation. DG is supported by the NSERC Discovery
Grant program and by the Perimeter Institute for Theoretical Physics. Research at
Perimeter Institute is supported in part by the Government of Canada through the
Department of Innovation, Science and Economic Development Canada and by the
Province of Ontario through the Ministry of Colleges and Universities.

\pagebreak
\section{An Explicit Differential for the Bosonic Pentagon Identity} \label{sec:boson} \paragraph{} As discussed in the introduction, there are two versions of the pentagon identity, the fermionic version and the bosonic version. There are natural vector spaces which reproduce quantum dilogarithms along with products of dilogarithms upon taking a graded trace. As we will review extensively in Section \ref{sec:fermion}, the fermionic pentagon identity can be upgraded to a literal isomorphism of such vector spaces (in fact an isomorphism of algebras). Passing to the traces, this corresponds to the fact that all coefficients in the series expansion on either the side with two factors or the side with three in the pentagon identity are positive, and the equality holds term by term without cancellations. \paragraph{} The bosonic pentagon identity is in stark contrast to this. The identity instead relies on heavy cancellations between coefficients on the side with three factors. Again there are natural state spaces, graded by a cohomological degree, which reproduce the inverse dilogarithm upon taking characters. However, because of the cancellations required the categorification is more subtle. 

\paragraph{} ``Cancellations" of terms in traces often corresponds, at the cohomological level, to states whose cohomological degrees differ by a unit being paired up by a nilpotent differential. It is therefore natural to wonder if there's an appropriate differential that acts on the bosonic state spaces that implements such cancellations at the homological level. The purpose of this section is to construct one such differential by hand and to perform some basic checks on whether it leads to a categorification of the bosonic pentagon identity.

\paragraph{} The first order of business in categorifying the bosonic pentagon identity is to spell out the vector space that categorifies the (inverse) quantum dilogarithm \bea \Phi(X_{\gamma})^{-1} = \prod_{n=0}^{\infty}(1+\fq^{2n+1}X_{\gamma})^{-1}.\eea The vector space $\mathcal{C}_{\gamma}$ that categorifies $\Phi(X_{\gamma})$ will be graded by three quantum numbers: the gauge charge $g$ valued in integer multiples $n \gamma$ of the elementary gharge $\gamma \in \Gamma$, the ghost number $\text{gh}$ valued in integers, and the twisted spin quantum number $s$ valued in half-integers. The dilogarithm will be recovered by taking the graded character \bea \Phi(X_{\gamma}) = \text{Tr}_{\mathcal{C}_{\gamma}}\big((-1)^{\text{gh}} \fq^{2s} X_{g}\big).\eea We introduce a basic generating object $a_{\gamma}$ such that the quantum numbers of $a_{\gamma}$ are \bea (g,\text{gh},s) \big(a_{\gamma} \big) = \big(\gamma, 1, \frac12 \big).\eea We now introduce a formal variable $\del$ which carries quantum numbers \bea (g, \text{gh},s) \big(\del \big) = (0,0,1). \eea The \textit{categorified bosonic dilogarithm} is defined to be \bea \mathcal{C}_{\gamma} := \text{Sym}^*\big(\mathbb{C}[\mathbf{\del}] a_{\gamma} \big),\eea where $\text{Sym}^*(V)$ denotes the symmetric algebra of a vector space $V$. 

\paragraph{}The basic physical motivation for this choice is that $\Phi(X_{\gamma})^{-1}$ counts the contribution of a single hypermultiplet field to a space of boundary local operators in the HT twist. The elementary field in the HT twist has the same quantum numbers as $a_{\gamma}$ and descendants $\mathbf{\del}^k a_{\gamma}$ represent holomorphic derivatives of the elementary field. See Appendix \ref{httwist} for more details. 

\paragraph{} We can be more explicit as follows. Let \bea a_{\gamma}^{[k]} := \del^k a_{\gamma} \eea for a non-negative integer $k$. The quantum numbers of $a_{\gamma}^{[k]}$ are simply $(\gamma, 1, k+\frac12)$. The vector space $\mathcal{C}_{\gamma}$ is then simply the polynomial ring of $a_{\gamma}^{[0]}, a_{\gamma}^{[1]}, a_{\gamma}^{[2]}, \dots$ \bea \mathcal{C}_{\gamma} = \mathbb{C}[a_{\gamma}^{[0]}, a_{\gamma}^{[1]}, a_{\gamma}^{[2]}, \dots]\eea where all gradings are additive under multiplication. We stress that the $a_{\gamma}^{[i]}$ obey by bosonic statistics \bea a_{\gamma}^{[i]} a_{\gamma}^{[j]} = a_{\gamma}^{[j]} a_{\gamma}^{[i]}\eea even though they carry ghost number $+1$. This is because in the twisted theory the ghost number and the fermion number are distinct. It is important to note that the space $\mathcal{C}_{\gamma}$ carries the structure of a differential algebra\footnote{Not to be confused with the notion of a \textit{differential-graded} algebra. In particular $\del$ has nothing to do with a differential in the sense of chain complexes.}, meaning the operator $\del$ can be extended to act from the single particle states to the full space \bea \del: \mathcal{C}_{\gamma} \rightarrow \mathcal{C}_{\gamma}\eea via the Leibniz rule: \bea \del(a_{\gamma}^{[k]} a_{\gamma}^{[l]}) &=& \del a_{\gamma}^{[k]} a_{\gamma}^{[l]} + a_{\gamma}^{[k]} \del a_{\gamma}^{[l]} \\ &=& a_{\gamma}^{[k+1]} a_{\gamma}^{[l]} + a_{\gamma}^{[k]}  a_{\gamma}^{[l+1]}. \eea As $\mathcal{C}_{\gamma}$ factorizes as a tensor product of bosonic Fock spaces generated by powers of $\del^k a_{\gamma}$ with each factor having character \bea (1+\fq^{2k+1} X_{\gamma})^{-1} ,\eea the character of $\mathcal{C}_{\gamma}$ is given by the inverse quantum dilogarithm $\Phi(X_{\gamma})^{-1}$.

\paragraph{} In order to categorify wall-crossing formulas, we need to categorify the contribution of multiple quantum dilogarithms with dyonic charges which are not mutually local. As particles with such charges cannot be simultaneously described by fields, the ``polynomials in the fields'' description of local operators breaks down. We propose here an ad-hoc rule which mimics the non-commutative multiplication rule that the quantum torus variables obey. Given two vector spaces $\mathcal{C}_{\gamma}$ and $\mathcal{C}_{\gamma'}$ that categorify the quantum dilogarithms $\Phi(X_{\gamma})$ and $\Phi(X_{\gamma'})$ for some $\gamma, \gamma' \in \Gamma$ we introduce the \textit{twisted tensor product}  \bea  \mathcal{C}_{\gamma}\, \widehat{\otimes}  \,\,\mathcal{C}_{\gamma'}, \eea defined as follows. As a vector space $\mathcal{C}_{\gamma} \widehat{\otimes}  \,\,\mathcal{C}_{\gamma'}$ is identical to the usual tensor product $\mathcal{C}_{\gamma} \otimes \mathcal{C}_{\gamma'}$. However, the gradings in $\mathcal{C}_{\gamma} \, \widehat{\otimes} \, \, \mathcal{C}_{\gamma'}$ are not the standard ones inherited from the tensor product of vector spaces. Instead, we impose the following rule. Suppose $v_1 \in \mathcal{C}_{\gamma}$ is a state with definite charges $(\gamma_1, f_1,s_1)$  and $v_{2} \in \mathcal{C}_{\gamma'}$ is a state with charges $(\gamma_2, f_2, s_2)$. Then we let \bea g (v_1 \widehat{\otimes} \, v_2) &=& \gamma_1 + \gamma_2,\\ \text{gh} (v_1 \widehat{\otimes} \, v_2) &=&  f_1 + f_2, \\  s(v_1 \widehat{\otimes} \, v_2) &=&  s_1 + s_2 +\frac12  \langle \gamma_1,  \gamma_2 \rangle , \eea so that the ghost numbers and gauge charges are additive, but the spin gets an additional contribution from the symplectic pairing on $\Gamma$. We can extend this in a similar way to a twisted tensor product of arbitrarily many $\mathcal{C}_{\gamma}$'s. This is essentially a point-splitting regularization: each element in the twisted tensor product represents a collection of elementary boundary local operators separated 
along the topological direction. The shift in $s$ is just the Poynting vector contribution from separated pairs of dyonic sources.  

\paragraph{The Left Vector Space $\mathcal{A}_{\text{L}}$:} Suppose that $\gamma_1$ and $\gamma_2$ are two elements of $\Gamma$ such that $\langle \gamma_1, \gamma_2 \rangle = 1$. We consider the natural vector space that categorifies the left-hand-side of the pentagon identity \bea \mathcal{A}_L := \mathcal{C}_{\gamma_1} \, \widehat{\otimes}\,\, \mathcal{C}_{\gamma_2}. \eea Explicitly, we can enumerate a basis of states for $\mathcal{A}_{\text{L}}$ as follows. Letting $a_{\gamma_1}, a_{\gamma_2}$ be the basic one-particle states, the state \bea \prod_{i=1}^n \del^{k_i} a_{\gamma_1} \, \widehat{\otimes} \, \prod_{j=1}^m \del^{l_j} a_{\gamma_2},\eea has quantum numbers \bea (g, \text{gh}, s) = \Big(n\gamma_1 + m \gamma_2, \,\, n+m, \,\, \sum_{i=1}^n k_i +\sum_{j=1}^m l_j +1 + \frac{nm}{2} \Big) .\eea It is useful to decompose the states into gauge charges \bea \mathcal{A}_{\text{L}} = \oplus_{\gamma \in \Gamma} \mathcal{A}_{\text{L},\gamma} .\eea We describe the space $\mathcal{A}_{n\gamma_1 + m\gamma_2}$ for low $n+m$. Firstly, we have that $\mathcal{A}_{\text{L}, \gamma_1}$ has a basis of states given by $a_{\gamma_1}$ and its derivatives \bea \{a_{\gamma_1}, \del a_{\gamma_1}, \del^2 a_{\gamma_1}, \dots \}.\eea Similarly, $\mathcal{A}_{\text{L}, \gamma_2}$ consists of $a_{\gamma_2}$ and its derivatives \bea \{a_{\gamma_2}, \del a_{\gamma_2}, \del^2 a_{\gamma_2}, \dots \}.\eea The first non-trivial twisted product appears in $\mathcal{A}_{\gamma_1 + \gamma_2}$, which has a basis of states given by \bea \{\del^k a_{\gamma_1} \, \widehat{\otimes} \,\, \del^l a_{\gamma_2} \}_{k,l= 0, 1, 2, \dots} .\eea The state labeled by $(k,l)$ has quantum numbers $(\text{gh}, s) = (2, k+l + \frac{3}{2})$. The minimum spin in $\mathcal{A}_{\gamma_1 + \gamma_2} $ is $\frac{3}{2}$ being realized by the state $a_{\gamma_1} \,\widehat{\otimes} \,a_{\gamma_2}$ with no derivatives.  

\paragraph{The Right Vector Space $\mathcal{A}_{\text{R}}$:} Let us now consider the vector space that categorifies the left hand side of the pentagon identity \bea \mathcal{A}_{\text{R}} := \mathcal{C}_{\gamma_2} \, \, \widehat{\otimes} \,\, \mathcal{C}_{\gamma_2 + \gamma_1} \,\, \widehat{\otimes} \,\, \mathcal{C}_{\gamma_1}. \eea We now write the basic element of $\mathcal{C}_{\gamma}$ as $\alpha_{\gamma}$ to distinguish it from the $a_{\gamma}$ appearing in the left vector space $\mathcal{A}_{\text{L}}$. The right vector space $\mathcal{A}_{\text{R}}$ has a basis consisting of states of the type: \bea \label{generalstateright} \prod_{i=1}^n \del^{k_i} \alpha_{\gamma_2} \widehat{\otimes} \,\prod_{j=1}^m \del^{l_j} \alpha_{\gamma_1 + \gamma_2} \widehat{\otimes} \,\prod_{s=1}^p \del^{r_s} \alpha_{\gamma_1}.\eea The state written above has gauge charge \bea g = (n+m)\gamma_1 + (m+p) \gamma_2,\eea ghost number \bea \text{gh} = n+m+p\eea and spin degree \bea s = \sum_{i=1}^n k_i + \sum_{j=1}^m l_j + \sum_{s=1}^p r_s + \frac{3}{2} - \frac{1}{2}(nm + np + mp).\eea Let's again look at the subspaces of low gauge charge. We have that $\mathcal{A}_{\text{R}, \gamma_1}$ and $\mathcal{A}_{\text{R}, \gamma_2}$ have a basis of states consisting of $\alpha_{\gamma_1}$ and their derivatives, and $\alpha_{\gamma_2}$ and their derivatives, respectively. The structure of $\mathcal{A}_{\text{R}, \gamma_1 + \gamma_2}$ is slightly more intricate. It consists of two summands. There is a summand with $\text{gh}=1$ consisting of states of the form \bea \{\alpha_{\gamma_1+ \gamma_2}, \del \alpha_{\gamma_1 + \gamma_2}, \del^2 \alpha_{\gamma_1 + \gamma_2}, \dots \},\eea the spin of $\del^{k} \alpha_{\gamma_1 + \gamma_2}$ being $k+ \frac{1}{2}$ as usual. There is another summand with $\text{gh}=2$ consisting of states which are non-trivial products of the form \bea \{ \del^k \alpha_{\gamma_2} \widehat{\otimes} \, \del^{l} \alpha_{\gamma_1} \}, \eea where the spin of the state labeled by the pair $(k,l)$ is now $k + l + \frac{1}{2}$. We notice that the minimum spin in $\mathcal{A}_{\text{R}, \gamma_1+ \gamma_2}$ is $\frac{1}{2}$ with two corresponding states given by $\alpha_{\gamma_1 + \gamma_2}$ and $\alpha_{\gamma_2}\, \widehat{\otimes} \, \alpha_{\gamma_1}$. 

\paragraph{} We notice that $\mathcal{A}_{\text{L}}$ and $\mathcal{A}_{\text{R}}$ are not isomorphic as vector spaces graded by the three quantum numbers $(g, \text{gh}, s)$. Indeed, we observed that the simplest mismatch occurs in the sector with gauge charge $\gamma_1 + \gamma_2$. We had that the minimum spin appearing in $\mathcal{A}_{\text{L}, \gamma_1 + \gamma_2}$ was $s= \frac{3}{2}$, in particular \bea \mathcal{A}_{\text{L}, \gamma_1 + \gamma_2, s=\frac{1}{2}} = \{0 \}.\eea On the other hand \bea \mathcal{A}_{\text{R}, \gamma_1 + \gamma_2, s = \frac{1}{2}} = \mathbb{C}^{\text{gh}=1} \oplus \mathbb{C}^{\text{gh}=2},\eea where $\mathbb{C}^{\text{gh}=1}$ and $\mathbb{C}^{\text{gh}=2}$ are the one-dimensional vector spaces with $\text{gh} = 1$ and $\text{gh}= 2$ respectively. The corresponding states are given by $\alpha_{\gamma_1+ \gamma_2}$ and $\alpha_{\gamma_2} \widehat{\otimes} \, \alpha_{\gamma_1}$. Therefore $\mathcal{A}_{\text{L}}$ and $\mathcal{A}_{\text{R}}$ are not isomorphic as graded vector spaces. 

\paragraph{} We now show how to equip $\mathcal{A}_{\text{L}}$ and $\mathcal{A}_{\text{R}}$ with operators \bea Q_{\text{L}}: \mathcal{A}_{\text{L}} \rightarrow \mathcal{A}_{\text{L}}, \,\,\,\,\,\,\,\, Q_{\text{R}}: \mathcal{A}_{\text{R}} \rightarrow \mathcal{A}_{\text{R}},\eea which preserve the gauge charge and spin, but increase the ghost number $\text{gh}$ by $+1$, which in addition satisfy \bea Q_{\text{L}}^2 =0, \,\,\, Q_{\text{R}}^2 =  0 .\eea In other words, we equip $\mathcal{A}_{\text{L}}$ and $\mathcal{A}_{\text{R}}$ with differentials $Q_{\text{L}}$ and $Q_{\text{R}}$ so these spaces now become chain complexes with the cohomological grading being given by the ghost number. We will then claim that with the appropriate differentials, the left and right chain complexes are homotopy equivalent \bea (\mathcal{A}_{\text{L}}, Q_{\text{L}}) \simeq (\mathcal{A}_{\text{R}}, Q_{\text{R}}).\eea The latter statement is a categorification of the bosonic pentagon identity. 

\paragraph{The Differential $Q_{\text{L}}$:} Since we are imposing that the differential preserve the gauge charge, we study the structure of the space $\mathcal{A}_{\text{L}, n\gamma_1 + m \gamma_2}$ of a fixed gauge charge. Recall that this space is constructed by acting arbitrarily many derivatives on a product of $n$ $a_{\gamma_1}$'s and a product of $m$ $a_{\gamma_2}$'s. In particular this means that the entire space $\mathcal{A}_{\text{L}, n\gamma_1 + m \gamma_2}$ lives in a fixed ghost number $\text{gh} = n + m$. The requirement that the differential \bea Q_{\text{L},n\gamma_1 + m\gamma_2} : \mathcal{A}_{\text{L}, n\gamma_1 + m \gamma_2} \rightarrow \mathcal{A}_{\text{L}, n\gamma_1 + m\gamma_2} \eea raise the ghost number by a unit then uniquely fixes it to vanish, \bea Q_{\text{L}} = 0.\eea 

\paragraph{The Differential $Q_{\text{R}}$:} We now come to the first main result of this paper, which is the construction of an explicit non-trivial differential \bea Q_{\text{R}}: \mathcal{A}_{\text{R}} \rightarrow \mathcal{A}_{\text{R}} .\eea As with the differential on the left complex, we study the possible differentials on the sectors of fixed gauge charge. First notice that $\mathcal{A}_{\text{R},\gamma_1}$ and $\mathcal{A}_{\text{R},\gamma_2}$ are both concentrated in a single ghost number $\text{gh}=1$ and so we must have \bea Q_{\text{R}} (\del^k \alpha_{\gamma_1}) = Q_{\text{R}}(\del^l \alpha_{\gamma_2}) = 0,\eea for all non-negative integers $k$ and $l$. The lowest gauge charge where $Q_{\text{R}}$ acts non-trivially is the sector $\mathcal{A}_{\text{R},\gamma_1 + \gamma_2}$. Since $Q_{\text{R}}$ also preserves the spin we can restrict our attention to a particular spin degree. Recall that the minimum spin degree in $\mathcal{A}_{\text{R}, \gamma_1 + \gamma_2}$ was $s= \frac{1}{2}$ which had two states, $\alpha_{\gamma_1 + \gamma_2}$ with $\text{gh}=1$ and $\alpha_{\gamma_2} \widehat{\otimes} \, \alpha_{\gamma_1}$ with $\text{gh}=2$. The ghost numbers of these two states differ precisely by a unit, and allow a non-trivial differential mapping one state to the other. Accordingly, we propose \bea \label{1to2differential} Q_{\text{R}}(\alpha_{\gamma_1 + \gamma_2}) = \alpha_{\gamma_2} \,\widehat{ \otimes} \, \alpha_{\gamma_1}.\eea The equation above along with a simple compatibility rule for derivatives actually determines the differential on the whole space $\mathcal{A}_{\text{L}, \gamma_1+ \gamma_2}$. Recall that the space $\mathcal{C}_{\gamma}$ is a $\mathbb{C}[\del]$-module, and naturally, so is the twisted tensor product $\mathcal{C}_{\gamma} \,\widehat{\otimes}\,\, \mathcal{C}_{\gamma'}$. The action of $\del$ on a twisted tensor product of states is simply given by the Leibniz rule \bea \del( v\,\widehat{\otimes} \, w) = \del v  \,\widehat{\otimes} \, w + v\, \widehat{\otimes } \, \del w.  \eea We then extend the differential \eqref{1to2differential} acting on states of gauge charge $\gamma_1 + \gamma_2$ and minimum spin $s = \frac{1}{2}$ to states of the same gauge charge $\gamma_1+ \gamma_2$ and arbitrary spin simply be requiring that $Q_{\text{R}}$ commute with $\del$ so that \bea Q_{\text{R}} (\del^{n} \alpha_{\gamma_1+ \gamma_2}) = \del^n \big(\alpha_{\gamma_2} \widehat{\otimes} \, \alpha_{\gamma_1} \big).\eea Thus for instance, \bea Q_{\text{R}} (\del \alpha_{\gamma_1 + \gamma_2} ) &=& \del \alpha_{\gamma_2} \widehat{\otimes } \, \alpha_{\gamma_1} + \alpha_{\gamma_{2} } \widehat{\otimes} \, \del \alpha_{\gamma_1}, \\   Q_{\text{R}} (\del^2 \alpha_{\gamma_1 + \gamma_2} ) &=& \del^2 \alpha_{\gamma_2} \widehat{\otimes } \, \alpha_{\gamma_1} +2\del \alpha_{\gamma_{2} } \widehat{\otimes} \, \del \alpha_{\gamma_1} +  \alpha_{\gamma_{2} } \widehat{\otimes} \, \del^2 \alpha_{\gamma_1}.\eea For arbitrary $n$ then we have \bea \label{diffonder} Q_{\text{R}}(\del^n \alpha_{\gamma_1+ \gamma_2} ) = \sum_{k=0}^n \binom{n}{k} \del^{n-k} \alpha_{\gamma_2} \widehat{\otimes} \, \del^k \alpha_{\gamma_1}.\eea This expression then defines a two-term complex \bea \begin{CD} 0 @>>>  \mathcal{A}^1_{\text{R}, \gamma_1+ \gamma_2} @>Q_{\text{R}}>> \mathcal{A}^2_{\text{R}, \gamma_1+ \gamma_2} @>>> 0.\end{CD} \eea We now extend the definition of $Q_{\text{R}}$ to higher gauge charge. The next step is to write the differential on states of gauge charge $2\gamma_1 + \gamma_2$ and $\gamma_1 + 2\gamma_2$. The states in these vector spaces are given simply by attaching on an $\alpha_{\gamma_2}$ to the left (for $\gamma_1 + 2\gamma_2$) or an $\alpha_{\gamma_1}$ to the right (for $2\gamma_1 + \gamma_2$) of a state of gauge charge $\gamma_1 + \gamma_2$. The differential on these states is defined by the previous formula with the $\alpha_{\gamma_1}$ and $\alpha_{\gamma_2}$ going along for the ride: \bea Q_{\text{R}}(\alpha_{\gamma_2} \widehat{\otimes} \alpha_{\gamma_1+ \gamma_2} ) &=& \alpha_{\gamma_2}^2 \widehat{\otimes} \, \alpha_{\gamma_1}, \\ Q_{\text{R}}(\alpha_{\gamma_1 + \gamma_2} \widehat{\otimes } \,\alpha_{\gamma_1}) &=& \alpha_{\gamma_1} \widehat{\otimes} \, \alpha_{\gamma_2}^2, \eea with natural generalizations when $\del$'s are involved.  The next non-trivial action of the differential is on states of gauge charge $2\gamma_1 + 2\gamma_2$. The space $ \mathcal{A}_{2\gamma_1 + 2\gamma_2}$ decomposes into states of ghost number $2,3$ and $4$. Accordingly, we will have a three-term complex for each spin $s=1,2,3,\dots$ \bea \begin{CD} 0 @>>> \mathcal{A}^{2}_{\text{R}, 2\gamma_1 + 2\gamma_2, s} @>Q_{\text{R}}>>  \mathcal{A}^{3}_{\text{R}, 2\gamma_1 + 2\gamma_2, s} @>Q_{\text{R}}>> \mathcal{A}^{4}_{\text{R}, 2\gamma_1 + 2\gamma_2, s} @>>>0 .\end{CD}\eea There is also a two-term complex in spin $s=0$ with a one-dimensional spaces in $\text{gh}=3,4$ where the $\text{gh}=3$ state is given by $\alpha_{\gamma_2} \widehat{\otimes} \, \alpha_{\gamma_1 + \gamma_2} \widehat{\otimes} \, \alpha_{\gamma_1}$ and the $\text{gh}=4$ state is given by $\alpha_{\gamma_2}^2 \widehat{\otimes} \, \alpha_{\gamma_1}^2 $. The differential pairs up the two $s=0$ states \bea Q_{\text{R}}(\alpha_{\gamma_2} \widehat{\otimes} \, \alpha_{\gamma_1 + \gamma_2} \widehat{\otimes} \, \alpha_{\gamma_1}) = \alpha_{\gamma_2}^2 \widehat{\otimes} \, \alpha_{\gamma_1}^2 , \eea consistent with the rule that the $\alpha_{\gamma_1}$ and $\alpha_{\gamma_2}$ go along for the ride. We now look at our first three-term complex by considering $s=1$. It is of the form \bea \begin{CD} 0 @>>> \mathbb{C} @>Q_{\text{R}}>> \mathbb{C}^3 @>Q_{\text{R}}>> \mathbb{C}^2 @>>>0,\end{CD}\eea where the non-trivial vector spaces are in ghost numbers $2,3$ and $4$ respectively. The state in $\text{gh}=2$ is given by \bea \alpha_{\gamma_1+ \gamma_2}^2.\eea The three states in $\text{gh}=3$ are given by \bea \{\del\alpha_{\gamma_2} \widehat{\otimes} \, \alpha_{\gamma_1 + \gamma_2} \widehat{\otimes} \, \alpha_{\gamma_1} \, , \,  \alpha_{\gamma_2} \widehat{\otimes} \, \del \alpha_{\gamma_1 + \gamma_2} \widehat{\otimes} \, \alpha_{\gamma_1} \, , \,  \alpha_{\gamma_2} \widehat{\otimes} \, \alpha_{\gamma_1 + \gamma_2} \widehat{\otimes} \, \del\alpha_{\gamma_1} \}.\eea The two states with $\text{gh}=4$ and $s=1$ are given by \bea \{ \alpha_{\gamma_2}\del \alpha_{\gamma_2} \widehat{ \otimes}\, \alpha_{\gamma_1}^2 \, , \,   \alpha_{\gamma_2}^2 \widehat{\otimes} \, \alpha_{\gamma_1} \del \alpha_{\gamma_1} \} . \eea We first specify the differential \bea \begin{CD}\mathbb{C}^3 @>Q_{\text{R}}>> \mathbb{C}^2 \end{CD} .\eea It is given by \bea Q_{\text{R}}(\del\alpha_{\gamma_2} \widehat{\otimes} \, \alpha_{\gamma_1 + \gamma_2} \widehat{\otimes} \, \alpha_{\gamma_1}) &=& \alpha_{\gamma_2}  \del\alpha_{\gamma_2} \widehat{\otimes} \, \alpha_{\gamma_1}^2, \\  Q_{\text{R}}(\alpha_{\gamma_2} \widehat{\otimes} \, \del\alpha_{\gamma_1 + \gamma_2} \widehat{\otimes} \, \alpha_{\gamma_1}) &=& \alpha_{\gamma_2}  \del\alpha_{\gamma_2} \widehat{\otimes} \, \alpha_{\gamma_1}^2 + \alpha_{\gamma_2}^2  \widehat{\otimes} \, \alpha_{\gamma_1}\del \alpha_{\gamma_1}, \\ Q_{\text{R}}( \alpha_{\gamma_2} \widehat{\otimes} \, \alpha_{\gamma_1 + \gamma_2} \widehat{\otimes} \, \del\alpha_{\gamma_1}) &=& \alpha_{\gamma_2}^2  \widehat{\otimes} \, \alpha_{\gamma_1}\del \alpha_{\gamma_1}. \eea This specifies the differential $Q_{\text{R}, s=1, \text{gh}=3, g=2\gamma_1 + 2\gamma_2}$. There is also a one-dimensional space in $\text{gh}=2$ spanned by the state \bea \{\alpha_{\gamma_1+ \gamma_2}^2 \},\eea and we would like to determine the action of the differential $Q_{\text{R}}$ on this. It will be determined by the requirement that the three-term complex with $s=1$ must be acylic. This uniquely fixes then up to a scale, the differential from the $\text{gh}=2$ state to $\text{gh}=3$ \bea \begin{CD} \mathbb{C} @>Q_{\text{R}}>> \mathbb{C}^3 ,\end{CD}\eea by the requirement that its image space coincide with the kernel of the map from $\text{gh}=3$ states to $\text{gh}=4$. The kernel of $Q_{\text{R}}|_{s=1,\text{gh}=3}$ is the one-dimensional space spanned by the state \bea \del\alpha_{\gamma_2} \widehat{\otimes} \, \alpha_{\gamma_1 + \gamma_2} \widehat{\otimes} \, \alpha_{\gamma_1} -  \alpha_{\gamma_2} \widehat{\otimes} \, \del \alpha_{\gamma_1 + \gamma_2} \widehat{\otimes} \, \alpha_{\gamma_1} + \alpha_{\gamma_2} \widehat{\otimes} \, \alpha_{\gamma_1 + \gamma_2} \widehat{\otimes} \, \del\alpha_{\gamma_1} .\eea Accordingly, we set \bea \label{gaugecharge22} Q_{\text{R}}(\alpha_{\gamma_1+\gamma_2}^2) = c(\del\alpha_{\gamma_2} \widehat{\otimes} \, \alpha_{\gamma_1 + \gamma_2} \widehat{\otimes} \, \alpha_{\gamma_1} -  \alpha_{\gamma_2} \widehat{\otimes} \, \del \alpha_{\gamma_1 + \gamma_2} \widehat{\otimes} \, \alpha_{\gamma_1} + \alpha_{\gamma_2} \widehat{\otimes} \, \alpha_{\gamma_1 + \gamma_2} \widehat{\otimes} \, \del\alpha_{\gamma_1}) \eea for some non-zero coefficient $c$.  With this then we have specified the differential $Q_{\text{R}}$ on states of gauge charge $2\gamma_1 + 2\gamma_2$ and spin $s=0$ and $s=1$. It is not immediate however, how to extend $Q_{\text{R}}$ on states of the same gauge charge, but of arbitrary spin. The Leibniz rule only allows us to deduce the action of $Q_{\text{R}}$ on total derivatives, for instance at states of the type $\del^k(\alpha_{\gamma_1+ \gamma_2}^2)$ in $\text{gh}=2$. However, not all states of higher spin are total derivatives. The Leibniz rule does not determine $Q_{\text{R}}(\del^{2}\alpha_{\gamma_1+ \gamma_2} \alpha_{\gamma_1+ \gamma_2})$ for instance. 

\paragraph{} This can be efficiently dealt with by the introduction of generating functions. Let \bea \alpha_{\gamma}(z) := \sum_{k=0}^{\infty} \frac{1}{k!} \del^k \alpha_{\gamma} \,z^k \in \mathcal{C}_{\gamma} \otimes \mathbb{C}[[z]], \eea where $\gamma$ can be any of $\gamma_1, \gamma_2$ or $\gamma_{1} + \gamma_{2}$. The space $\mathcal{C}_{\gamma}$ that categorifies the dilogarithm can be recovered from products of generating functions in a standard way: the state \bea  \del^{k_1} \alpha_{\gamma} \dots \del^{k_n} \alpha_{\gamma}\eea can be recovered from the product of generating functions \bea \alpha_{\gamma}(z_1) \dots \alpha_{\gamma}(z_n) \in \mathcal{C}_{\gamma}[[z_1, \dots, z_n]],\eea by applying the differential operator \bea \del^{k_1}_{z_1} \dots \del^{k_n}_{z_n}\eea to it, and evaluating it at $(z_1, \dots, z_k) = 0$. In a similar way, one can recover all states in the twisted tensor products of categorified dilogarithms from twisted products of generating functions \bea \prod_{i=1}^m \alpha_{\gamma_2}(z_i) \,\widehat{\otimes} \, \prod_{j=1}^n \alpha_{\gamma_{1}+ \gamma_2}(w_j) \,\widehat{\otimes} \, \prod_{s=1}^p\alpha_{\gamma_1}(y_s).\eea By assigning the monomial $z^k$ spin quantum number $-k$, the generating function $\alpha_{\gamma}(z)$ has spin $\frac{1}{2}$. 

\paragraph{} Let's then reformulate what we know about our differential in terms of the formal generating power series. The differential \bea \label{1to2withz} Q_{\text{R}}\big(\alpha_{\gamma_1+ \gamma_2}(z) \big) = \alpha_{\gamma_2}(z) \, \widehat{\otimes} \, \alpha_{\gamma_1}(z)  \eea captures the action of $Q_{\text{R}}$ on all states of gauge charge $\gamma_1 + \gamma_2$, reproducing \eqref{diffonder}.  Next comes the action of states of gauge charge $2\gamma_1 + \gamma_2$ and $2\gamma_2 + \gamma_1$ which is defined in a straightforward way from \eqref{1to2withz} where additional $\alpha_{\gamma_1}(z)$ or $\alpha_{\gamma_2}(z)$ go along for the ride. For a while now we will focus on diagonal gauge charges $n\gamma_1 + n\gamma_2$ only. The next non-trivial differential comes about when $n=2$. We choose \bea \label{diff22z} \begin{split} Q_{\text{R}}\big(\alpha_{\gamma_1+ \gamma_2}(z) \alpha_{\gamma_1 + \gamma_2}(w) \big) = \,\,\,\,\,\,\,\,\,\,\,\,\,\,\,\,\,\,\,\,\,\,\,\,\,\,\,\,\,\,\,\,\,\,\,\,\,\,\,\,\,\,\,\,\,\,\,\,\,\,\,\,\,\,\,\,\,\,\,\,\,\,\,\,\,\,\,\,\,\,\,\,\,\,\,\,\,\,\,\,\,\,\,\,\,\,\,\,\,\,\,\,\,\,\,\,\,\,\,\,\,\,\,\,\,\,\,\,\,\,\,\,\,\,\,\,\,\,\,\,\,\,\,\,\,\,\,\,\,\,\,\,\,\,\,\,\,\,\,\,\,\, \\ \del_{z} \alpha_{\gamma_2} (z) \,\widehat{\otimes} \, \alpha_{\gamma_1 + \gamma_2}(w) \, \widehat{\otimes} \, \alpha_{\gamma_1}(z) -   \alpha_{\gamma_2} (z) \,\widehat{\otimes} \, \del_{w}\alpha_{\gamma_1 + \gamma_2}(w) \, \widehat{\otimes} \, \alpha_{\gamma_1}(z)   \\ + \alpha_{\gamma_2} (z) \,\widehat{\otimes} \, \alpha_{\gamma_1 + \gamma_2}(w) \, \widehat{\otimes} \, \del_{z}\alpha_{\gamma_1}(z) +  \del_{w} \alpha_{\gamma_2} (w) \,\widehat{\otimes} \, \alpha_{\gamma_1 + \gamma_2}(z) \, \widehat{\otimes} \, \alpha_{\gamma_1}(w)  \\ -   \alpha_{\gamma_2} (w) \,\widehat{\otimes} \, \del_{z}\alpha_{\gamma_1 + \gamma_2}(z) \, \widehat{\otimes} \, \alpha_{\gamma_1}(w)  +  \alpha_{\gamma_2} (w) \,\widehat{\otimes} \, \alpha_{\gamma_1 + \gamma_2}(z) \, \widehat{\otimes} \, \del_{w}\alpha_{\gamma_1}(w)  .\end{split}\eea Specializing to $z=w=0$ reproduces the equation \eqref{gaugecharge22} with $c=2$. In addition it also determines the action of $Q_{\text{R}}$ on all states of gauge charge $2\gamma_1 + 2\gamma_2$ and arbitrary spin. Defining the differential for states of higher ghost number is also straightforward. On states of gauge charge $2\gamma_1 + 2\gamma_2$ and $\text{gh}=3$ we set \bea Q_{\text{R}} \big(\alpha_{\gamma_2}(z) \, \widehat{\otimes} \, \alpha_{\gamma_1+\gamma_2}(w) \,\widehat{\otimes} \, \alpha_{\gamma_1}(y)  \big) = \alpha_{\gamma_2}(z) \alpha_{\gamma_2}(w) \widehat{\otimes} \alpha_{\gamma_1}(w) \alpha_{\gamma_1}(y).\eea We have successfully defined a differential on the sector with gauge charge $2\gamma_1+2\gamma_2$ and arbitrary spin \bea Q_{\text{R}, 2\gamma_1 + 2\gamma_2} : \mathcal{A}_{\text{R},2\gamma_1 + 2\gamma_2} \rightarrow \mathcal{A}_{\text{R},2\gamma_1 + 2\gamma_2}  \eea provided that the map $Q_{\text{R}}$ squares to zero pointwise in $z,w$. It is a straightforward calculation to check that this is indeed the case. Applying $Q_{\text{R}}$ to \eqref{diff22z} we find that the first three terms produce four terms which cancel against the four terms produced by the last three terms. $Q_{\text{R}}^2=0$ due to a straightforward pairwise cancellation of terms.

\paragraph{} We would like to generalize \eqref{diff22z} to diagonal gauge charges $n\gamma_1 + n\gamma_2$ for arbitrary $n$. We notice that \eqref{diff22z} can be rewritten in a suggestive form. It can be rewritten as \bea \label{diff22znice}\begin{split} Q_{\text{R}}\big(\alpha_{\gamma_1+ \gamma_2}(z) \alpha_{\gamma_1 + \gamma_2}(w) \big) = \big(\del_{z} - \del_{w} \big) (\alpha_{\gamma_2}(z) \, \widehat{\otimes} \, \alpha_{\gamma_1 + \gamma_2}(w) \,\widehat{\otimes} \, \alpha_{\gamma_1}(z) ) \\ + \big(\del_{w} - \del_{z} \big) (\alpha_{\gamma_2}(w) \, \widehat{\otimes} \, \alpha_{\gamma_1 + \gamma_2}(z) \,\widehat{\otimes} \, \alpha_{\gamma_1}(w) )  .\end{split}\eea The next step is to try writing $Q_{\text{R}}$ for states of gauge charge $3\gamma_1+3\gamma_2$. We find that the expression \bea \begin{split} Q_{\text{R}} \big(\alpha_{\gamma_1+ \gamma_2}(z) \alpha_{\gamma_1 + \gamma_2} (w) \alpha_{\gamma_1 + \gamma_2}(y) \big) = \,\,\,\,\,\,\,\,\,\,\,\,\,\,\,\,\,\,\,\,\,\,\,\,\,\,\,\,\,\,\,\,\,\,\,\,\,\,\,\,\,\,\,\,\,\,\,\,\,\,\,\,\,\,\,\,\,\,\,\,\,\,\,\,\,\,\,\,\,\,\,\,\,\,\,\,\,\,\,\,\\ \big(\del_{z}-\del_{w}\big) \big(\del_{z}-\del_{y} \big) \big( \alpha_{\gamma_2}(z) \,\widehat{\otimes} \, \alpha_{\gamma_1+\gamma_2}(w) \alpha_{\gamma_1+ \gamma_2}(y) \,\widehat{\otimes} \, \alpha_{\gamma_1}(z) \big) \\ +\big(\del_{w}-\del_{z}\big) \big(\del_{w}-\del_{y} \big) \big( \alpha_{\gamma_2}(w) \,\widehat{\otimes} \, \alpha_{\gamma_1+\gamma_2}(y) \alpha_{\gamma_1+ \gamma_2}(z)  \,\widehat{\otimes} \, \alpha_{\gamma_1}(w) \big) \\ + \big(\del_{y}-\del_{z}\big) \big(\del_{y}-\del_{w} \big) \big( \alpha_{\gamma_2}(y) \,\widehat{\otimes} \, \alpha_{\gamma_1+\gamma_2}(z) \alpha_{\gamma_1+ \gamma_2}(w)  \,\widehat{\otimes} \, \alpha_{\gamma_1}(y) \big) , \end{split}\eea is a natural generalization of \eqref{diff22znice}. A short calculation shows that it also satisfies $Q^2_{\text{R}} = 0$. Encouraged by this, we are ready to formulate the general differential.  

\paragraph{Theorem} Let $ Q_{\text{R}} : \mathcal{A}_{\text{R}, n\gamma_1+n\gamma_2} \rightarrow  \mathcal{A}_{\text{R}, n\gamma_1+n\gamma_2} $ be defined via \begin{align} \label{pentagondifferential1} Q_{\text{R}}\big(\alpha_{\gamma_1+\gamma_2}(z_1) \dots \alpha_{\gamma_1+\gamma_2}(z_n) \big) = \sum_{i=1}^{n} \prod_{j\neq i} \big(\del_{z_i} - \del_{z_j} \big) \big(\alpha_{\gamma_{2}}(z_i) \widehat{\otimes} \, \prod_{j \neq i} \alpha_{\gamma_1+ \gamma_2}(z_j) \widehat{\otimes} \, \alpha_{\gamma_{1}}(z_i) \big), 
\end{align} and extended to a map acting on the full space $\mathcal{A}_{\text{R}}$ via \begin{align} Q_{\text{R}} \big( \prod_{i=1}^n \alpha_{\gamma_2}(z_i) \widehat{\otimes} \, \prod_{j=1}^m \alpha_{\gamma_1 + \gamma_2}(w_j) \widehat{\otimes} \,\prod_{k=1}^p \alpha_{\gamma_1}(y_k) \big) =    \prod_{i=1}^n \alpha_{\gamma_1}(z_i) \widehat{\otimes} \, Q_{\text{R}}\big(\prod_{j=1}^m \alpha_{\gamma_1 + \gamma_2}(w_j)  \big) \widehat{\otimes} \prod_{k=1}^p \alpha_{\gamma_1}(y_k) .
\end{align} Then $Q_{\text{R}} \circ Q_{\text{R}} =0$.

\begin{proof} $Q_{\text{R}}^2\big( \alpha_{\gamma_1+ \gamma_2}(z_1) \dots \alpha_{\gamma_1+\gamma_2}(z_n) \big)$ can be organized as a sum over ordered pairs $(i,j)$ of distinct elements in $\{1,\dots, n\}$ as follows \bea \begin{split} Q_{\text{R}}^2 \Big( \alpha_{\gamma_1+ \gamma_2}(z_1) \dots \alpha_{\gamma_1+\gamma_2}(z_n) \Big) = \,\,\,\,\,\,\,\,\,\,\,\,\,\,\,\,\,\,\,\,\,\,\,\,\,\,\,\,\,\,\,\,\,\,\,\,\,\,\,\,\,\,\,\,\,\,\,\,\,\,\,\,\,\,\,\,\,\,\,\,\,\,\,\,\,\,\,\,\,\,\,\,\,\,\,\,\,\,\,\,\,\,\,\,\,\,\,\,\,\,\,\,\,\,\,\,\,\,\,\,\,\,\,\,\,\,\,\,\,\,\,\,\,\,\,\,\,\,\,\,\,\,\,\,\,\,\,\,\,\,\,\, \\ \sum_{ i\neq j} ( \del_{i} - \del_{j} ) \prod_{\substack{a\neq i \\ a\neq j}} (\del_{i} - \del_{a}) (\del_{j} - \del_{a}) \big(\alpha_{\gamma_2}(z_i) \alpha_{\gamma_2}(z_j) \,\widehat{\otimes} \, \prod_{\substack{a\neq i \\ a \neq j}} \alpha_{\gamma_1+\gamma_2}(z_a) \,\widehat{\otimes} \, \alpha_{\gamma_1}(z_i) \alpha_{\gamma_2}(z_j) \big) .  \end{split}\eea The terms in the sum cancel in pairs because the term corresponding to the ordered pair $(i,j)$ is precisely the negative of the term corresponding to the ordered pair $(j,i)$, leaving us with \bea Q_{\text{R}}^2 \Big( \alpha_{\gamma_1+ \gamma_2}(z_1) \dots \alpha_{\gamma_1+\gamma_2}(z_n) \Big) = 0.\eea
\end{proof} We have therefore completely specified a differential on the vector space $\mathcal{A}_{\text{R}}$, making it into a triply graded chain complex $(\mathcal{A}_{\text{R}}, Q_{\text{R}})$. We claim that this chain complex is homotopy equivalent to the left complex $(\mathcal{A}_{\text{L}}, Q_{\text{L}} = 0)$. One can make some basic checks by directly working out the cohomology of the right complex for low enough spin and gauge charge. However the complexes quickly become large and somewhat unmanageable to work with by hand\footnote{It is an interesting problem to directly compute, using techniques from homological algebra, the cohomology of the right complex $\mathcal{A}_{\text{R}}$ either with respect to the differential $Q_{\text{R}}$ written in \eqref{pentagondifferential1} or with respect to the differential $Q_{\text{K}}$ to be introduced later in the text \eqref{qkoszul}.}. We are able to use \textsc{Mathematica} to check that the Poincare series for the cohomology of the right chain complex $(\mathcal{A}_{\text{R}}, Q_{\text{R}})$ agrees with the Poincare series of the left complex $\mathcal{A}_{\text{L}}$ up to gauge charge $3\gamma_1 + 3\gamma_2$, and spin $s=8$.

\paragraph{} In the rest of this paper we will prove this claim by applying Koszul duality to the cohomological Hall algebra of the $A_{2}$ quiver. In fact Koszul duality will buy us a stronger result which upgrades the quasi-isomorphism of chain complexes to a quasi-isomorphism of differential graded algebras. Letting the associative products on either side be $\cdot_{\text{L}}$ and $\cdot_{\text{R}}$, we will therefore arrive at the \textit{categorified} bosonic pentagon identity \bea (\mathcal{C}_{\gamma_1} \,\widehat{\otimes} \,\,\mathcal{C}_{\gamma_2} \, , \, Q_{\text{L}} = 0, \cdot_{\text{L}}) \simeq (\mathcal{C}_{\gamma_2}\, \widehat{\otimes}\,\, \mathcal{C}_{\gamma_1+ \gamma_2}\, \widehat{\otimes} \,\, \mathcal{C}_{\gamma_1} \, ,\, Q_{\text{R}} , \cdot_{\text{R}}).\eea

\pagebreak

\section{The Fermionic Pentagon Identity and COHA} \label{sec:fermion}
\paragraph{} There is a beautiful way to prove and simultaneously categorify the fermionic pentagon identity \bea \Phi(X_{\gamma_2}) \Phi(X_{\gamma_1}) = \Phi(X_{\gamma_1}) \Phi(X_{\gamma_1+ \gamma_2}) \Phi(X_{\gamma_2})\eea by using the theory of cohomological Hall algebras (conventionally abbreviated to COHA) developed by Kontsevich and Soibelman \cite{Kontsevich:2010px}.

\paragraph{} Roughy speaking, the Cohomological Hall Algebra $\mathcal{H}(Q)$ is defined as a vector space as the cohomology of a space of representations of some quiver $Q$. This definition essentially identifies $\mathcal{H}(Q)$ with the space of supersymmetric states for a supersymmetric quantum mechanics modelled on $Q$. It is thus a natural object to consider in 
any situation in which the BPS particles of the 4d theory can be described by such a quiver quantum mechanics. This includes many situations where the four-dimensional theory can be embedded in string theory with BPS particles arising as bound states of BPS branes. The algebra structure on $\mathcal{H}(Q)$ can also be interpreted as encoding how a bound state can be split into two. 
The main claim of our companion paper \cite{future} is that this encoding is a form of Koszul duality. 

\paragraph{} The COHA relevant to the pentagon identity is the COHA for the $A_2$ quiver, the quiver with two nodes and one directed arrow between them. Denoting this algebra as $\mathcal{H}(A_2)$, it can be explicitly described using generators and relations. There are two sets of generators \bea \{e^1_n \}_{n=0,1,2,\dots}, \,\,\,\,\, \{e^2_{n}\}_{n=0,1,2, \dots}\eea corresponding to the two nodes of the quiver, where \bea (\text{gh}, g, s)(e^a_n) = \left(0, \gamma_a, n+\frac{1}{2}\right), \,\,\, a=1,2.\eea These generators are subject to relations which firstly state that these the generators corresponding to each node individually form an infinite Grassmann algebra \bea e^1_n e^1_m + e^1_m e^1_n = 0, \\ e^2_n e^2_m + e^2_m e^2_n = 0,\eea along with a fundamental relation that describes how to multiply two generators of different kind when they are in the ``wrong" order. This relation states that \bea \label{cohaexchange} e^1_n e^2_m = e^2_{m+1} e^1_n - e^2_m e^1_{n+1}. \eea The spin degrees in the COHA again employ the twisted tensor product rule so that the spin of both sides in the equation is indeed \bea \left(n+ \frac{1}{2}\right) + \left(m + \frac{1}{2}\right) + \frac{1}{2}. \eea

\paragraph{} The relations above allow us to write a Poincare-Birkhoff-Witt (PBW) basis for $\mathcal{H}(A_2)$: any element can be brought to a linear combination of elements of the form \bea e^2_{n_1} \dots e^2_{n_k} e^1_{m_1} \dots e^1_{m_l}.\eea Either of the two families of $e^i_n$ generates an exterior sub-algebra of $\mathcal{H}(A_2)$, which we can denote as $\cE_{\gamma_i}$. The PBW basis tells us that we can identify \bea \mathcal{H}(A_2) = \cE_{\gamma_2} \widehat \otimes \, \cE_{\gamma_1}. \eea The character of each subalgebra is a quantum dilogarithm. The natural graded trace of the COHA \bea \text{Tr}_{\mathcal{H}(A_2)} \big( (-1)^{\text{gh}} q^{s} X_{g} \big)\eea therefore coincides with \bea \Phi(X_{\gamma_2}) \Phi(X_{\gamma_1}).\eea 
Incidentally, $\cE_{\gamma}$ coincides with the COHA $\mathcal{H}(A_1)$ for a quiver with a single node and no loops, which is the BPS quiver for a single free hypermultiplet. 

\paragraph{}Remarkably there is an alternate PBW basis for $\mathcal{H}(A_2)$ (in fact there are two as we will remark later). If we let \bea e^{(12)}_n := e^2_0 e^1_n, \,\,\,\, n=0,1,2, \dots, \eea the original COHA relations imply that the collection $\{e^{(12)}_n \}_{n=0,1,2,\dots}$ also spans an infinite Grassmann algebra \bea \{ e^{(12)}_n, e^{(12)}_m \} = 0. \eea Moreover, straightforward computation allows one to deduce the following set of relations: \bea \label{rel1} e^2_{n} e^1_{m} &=& e^{(12)}_{n+m} + \sum_{k=1}^n e^1_{m+n - k} e^2_{k-1}, \\ \label{rel2} e^2_{n} e^{(12)}_{m} &=& -\sum_{k=1}^{n} e^{(12)}_{m+n-k} e^{2}_{k-1}, \\ \label{rel3} e^{(12)}_n e^1_m &=& \sum_{k=1}^{m} e^1_{m+n-k} e^{(12)}_{k-1} -\sum_{k=1}^{n} e^1_{m+n-k} e^{(12)}_{k-1}.\eea These relations provide an alternate PBW basis for $\mathcal{H}(A_2)$ consisting of elements of the form \bea e^{1}_{p_1} \dots e^1_{p_k} e^{(12)}_{q_1} \dots e^{(12)}_{q_l} e^{2}_{r_1} \dots e^{2}_{r_m}.\eea Thus we can decompose \bea \mathcal{H}(A_2) = \cE_{\gamma_1} \widehat \otimes \, \cE_{\gamma_1+ \gamma_2} \widehat \otimes  \, \cE_{\gamma_2}. \eea Equating the graded character computed from this alternate basis \bea \Phi(X_{\gamma_1}) \Phi(X_{\gamma_1+\gamma_2}) \Phi(X_{\gamma_2}) \eea to the one from the original basis gives us the fermionic pentagon identity. The categorification of the fermionic pentagon identity is therefore the statement that the two different presentations of the COHA indeed give the same underlying algebra $\mathcal{H}(A_2)$. 

\paragraph{} Curiously, there is an alternative definition \bea \tilde e^{(12)}_n := e^2_n e^1_0, \,\,\,\, n=0,1,2, \dots, \eea which also defines a fermionic sub-algebra and gives rise to an alternative ``three-particle" PBW basis. The physical significance of this alternative is obscure to us. Mathematically, the PBW bases should be associated to a Harder-Narasimhan filtration of quiver representations \cite{Kontsevich:2010px}. It would be instructive to derive and compare the filtrations associated to the two alternative bases. 

\paragraph{} We can define generating functions \bea e^i(z) \equiv \sum_{n=0}^\infty \frac{e^i_n}{z^{n+1}}\eea
It is straightforward to prove that the translation $e^i(z) \to e^i(z+w)$, where the right hand side is expanded in positive powers of $w$, is an algebra automorphism $\mathcal{H}(A_2) \to \mathcal{H}(A_2)[[w]]$ which is compatible with the above decompositions and sub-algebras. 
\pagebreak

\section{Interlude on Koszul Duality} \label{sec:koszul} Our strategy to deduce the categorification of the bosonic pentagon identity will be to compute the Koszul duals of the two different presentations of the COHA $\mathcal{H}(A_2)$. We use this section as an opportunity to summarize this key notion. 

\paragraph{} Koszul duality as discussed in the mathematics community is an idea of considerable age. Stemming from the world of homological algebra, mathematicians have indeed long appreciated its power to give insights into a wide array of subjects ranging from algebraic geometry, to deformation theory and homotopy theory. For a mathematical account of these developments, we refer the reader to the set of lecture notes \cite{Brantner} and references therein. Only in the past decade has Koszul duality began to play a significant role in theoretical physics, largely due to the work of Kevin Costello\footnote{We apologize in advance to other early adopters whose work we may not be familiar with.} \cite{Costello:2013zra,Costello:2017fbo, Costello:2020jbh}. Since it also plays a fundamental role in our work, we give a brief account. For another physically motivated account, see \cite{Paquette:2021cij}.

\subsection{Formal Aspects}
 
\paragraph{} Let $A$ be a graded, associative algebra equipped with an augmentation. An augmentation of $A$ is a homomorphism of algebras \bea \varepsilon: A \rightarrow \mathbb{C}\eea so that $\varepsilon$ gives $\mathbb{C}$ the structure of an $A$-module. The kernel of $\varepsilon$ is a two-sided ideal of $A$ known as the augmentation ideal and is denoted by $\bar A = \text{ker}(\varepsilon)$. 

\paragraph{} Koszul duality associates to $A$ a certain differential-graded algebra $A^!$ built from $\bar A$. Let \bea T^{\bullet}(\bar{A}^{\vee}[1]) = \mathbb{C} \oplus \bar{A}^{\vee}[1] \oplus \big(\bar{A}^{\vee}[1] \big)^{\otimes2} \oplus \dots,\eea be the tensor algebra of the dual of the degree-shifted\footnote{The notation $V[p]$ means we change the cohomological degree/ghost number of each homogeneous element by adding $p$ units, i.e an element of $V$ carrying degree $k$ is assigned degree $k+p$ in $V[p]$. } augmentation ideal. Then  $A^!$ is defined as the differential graded algebra \bea  \big(T^{\bullet}(\bar{A}^{\vee}[1]), Q \big) \eea with a differential $Q$ defined by using the product on $\bar{A}$ as follows. Let $\{e_a \}$ be a basis for $\bar{A}$, $m^a_{bc}$ the structure constants for the product \bea e_a \cdot e_b = m_{ab}^c e_c\eea and $\{e^a \}$ a basis for $\bar{A}^{\vee}[1]$. The differential is given by \bea Q e^a = m^a_{bc} \, e^b \otimes e^c\eea and extended to the higher tensor products by applying the Leibniz rule with respect to the product of words in the tensor algebra. $Q$ defines a nilpotent differential precisely due to the associativity of the product $m$. 

\paragraph{} The definition of the Koszul dual $A^{!}$ can be usefully formulated in terms of a universal property. Let $B$ be a differential graded algebra with differential $d_B$ and multiplication map encoded in terms of some structure constants \bea b_i b_j = (m_{B})_{ij}^k b_k\eea expressed in some basis $\{b_i\}$ for $B$. Consider $A\otimes B$ with the standard tensor product dg-algebra structure. Recall that a Maurer-Cartan element in a differential graded algebra with differential $d$ and multiplication $\cdot$ is a degree $+1$ element $\mu$ that satisfies the equation \bea d\mu = \mu \cdot \mu. \eea  The universal property states that the Koszul dual $A^{!}$ for an augmented algebra $A$ is the differential graded algebra such that for any differential-graded algebra $B$, Maurer-Cartan elements in the dg-algebra $A \otimes B$ coincide with dg-algebra homomorphisms $A^{!} \rightarrow B$. The dg-algebra $A^{!}$ as formulated above is easily seen to satisfy this property. The Maurer-Cartan equation for an element \bea \mu = \mu^{ai} e_a \otimes b_i \in A\otimes B\eea reads \bea \mu^{ai} e_a \otimes d_{B}(b_i) = \mu^{ai} \mu^{bj} (m_A)_{ab}^c (m_{B})_{ij}^k e_c \otimes b_k.\eea On the other hand a dg-algebra homomorphism $\phi:A^{!} \rightarrow B$ in particular determines constants $C^{ai}$ \bea \phi(e^a) = C^{ai} \, b_i.\eea Requiring that $\phi$ be an algebra homomorphism gives \bea \phi(e^a \otimes e^b) = C^{ai}C^{bj} (m_B)_{ij}^k b_k.\eea From this it's easy to see that further requiring $\phi$ to be a chain map \bea \phi(Q\, e^a) = d_B \, \phi(e^a)\eea precisely reproduces the Maurer-Cartan equation for the element $C = C^{ai}e_a \otimes b_j$.

\paragraph{} One can modify the definition of the Koszul dual so that it applies to not just graded augmented associative algebras, but to any augmented\footnote{An augmented $A_{\infty}$-algebra $A$ is one which comes equipped with an $A_{\infty}$-morphism from $A$ to $\mathbb{C}$. Equivalently $\mathbb{C}$ carries the structure of an $A_{\infty}$-module.} $A_{\infty}$-algebra $A$ subject to some finiteness conditions. Letting \bea m_k: \bar{A}^{\otimes k} \rightarrow \bar{A} \eea be the $k$th $A_{\infty}$-map, the differential on the Koszul dual dg-algebra $A^{!} = T^{\bullet} \bar{A}^{\vee}[1]$ is simply modified to take the form \bea Q e^a = (m_1)^a_b \, e^a + (m_2)^a_{bc} \, e^b \otimes e^c + (m_3)^a_{bcd}\, e^b \otimes e^c \otimes e^d + \dots \,.\eea It is nilpotent precisely due to the $A_{\infty}$-associativity relations. The universal property for Koszul duals still holds in this setting: if $B$ is any differential-graded algebra and $A$ is an $A_{\infty}$-algebra, then recall that $A\otimes B$ comes equipped with a natural $A_{\infty}$-structure. Denoting the $k$th $A_{\infty}$-map on $A\otimes B$ as $m_k^{A\otimes B}$, one can show that Maurer-Cartan elements, namely the elements $\mu \in A\otimes B$ that satisfy the $A_{\infty}$ Maurer-Cartan equation \bea m^{A\otimes B}_1 (\mu) = m^{A\otimes B}_2(\mu\otimes \mu) + m^{A\otimes B}_3(\mu \otimes \mu \otimes \mu) + \dots \eea are still in correspondence with dg-algebra homomorphisms $A^{!} \rightarrow B$. The generalization of $A^{!}$ to apply in the setting of dg-algebras in particular allows one to iterate the operation of taking the Koszul dual and define $(A^!)^!$.  If $A$ is sufficiently nice (aka Koszul), $(A^!)^!$ will be equivalent (quasi-isomorphic) to $A$ as an $A_{\infty}$-algebra. 

\paragraph{} A particularly nice situation is one where the dg algebra $A^!$ itself is equivalent to an actual algebra (as opposed to a more general $A_{\infty}$-algebra), which we will also denote as $A^!$. As we review below, this typically occurs when $A$ is defined by a collection of generators which satisfy quadratic relations. When $A$ is Koszul, $A^!$ is defined by a dual set of quadratic relations. The prototypical examples of such quadratic Koszul pairs are symmetric and exterior algebras $S^\bullet V^\vee[1]$ and $\bigwedge^\bullet V$, such as ${\mathcal C}_\gamma$ and $\cE_\gamma$. 

\paragraph{} In a typical physical setting, $A$ is an algebra of local operators along some topological line in the physical system. The augmentation represents some choice of vacuum or compactification which gives a one-dimensional space of states, so that local operators on the line can be assigned expectation values $A \to \bC$ compatible with the operator ring structure. Maurer-Cartan elements $\mu$ represent couplings of the system to an auxiliary one-dimensional system with operator algebra $B$, giving composite line defects which are compatible with the vacuum. The line defects then are classified by maps $A^! \to B$, so that $A^!$ is a sort of gauge algebra for the theory. 

\paragraph{} In some physical applications, calculations proceed in the opposite direction: information about the line defects themselves can be employed to derive a gauge algebra $A$ and then local operators can be computed as $A^!$ if a Koszul property holds. In our setup, the COHA $\cH(Q)$ should be analogous to a gauge algebra and $\cH(Q)^!$ to a space of local operators. We will explore this idea further in a companion paper \cite{future}.  

\paragraph{} For completeness, we also mention the algebraic origin for the construction\footnote{For further details see \cite{Weibel}.}: $A^!$ coincides with $\mathbf{R}\text{Hom}_A(\mathbb{C}, \mathbb{C})$, the space of derived endomorphisms of $\mathbb{C}$, viewed as an $A$-module via the augmentation $\varepsilon$. The notion of derived morphism between left $A$-modules $M$ and $N$ is a refinement of the space of $A$-equivariant linear maps $\text{Hom}_A(M,N)$ from $M$ to $N$. The way this is phrased in homological algebra is that we want to replace the functor \bea N \rightarrow \text{Hom}_A(M,N)\eea from the category of left $A$-modules to the category of complex vector spaces with its corresponding right derived functor.

\paragraph{} In more detail, suppose $P$ is chain complex  i.e a graded vector space $P = \oplus_{n\geq0} P_n$ with a degree $-1$ differential\footnote{Equivalently, flip the sign of gradations so that $P$ becomes concentrated in negative degrees and $d$ has degree $+1$. } $d:P \rightarrow P$ \bea \begin{CD} \dots @>d>> P_2  @>d>> P_1 @>d>> P_0 @>>> 0\end{CD}\eea which is also an $A$-module, and that moreover each $A$-module map is a chain map. $P$ is said to be a left resolution of $M$ if there is a map $u: P_0 \rightarrow M$ such that the sequence of maps \bea \begin{CD} \dots @>d>> P_1 @>d>> P_0 @>u>> M @>>> 0 \end{CD} \eea is an exact sequence. $P$ is said to be a projective resolution if the $A$-module $P$ is projective. One often writes a resolution $P$ of $M$ simply as $u: P \rightarrow M$. Having chosen a projective resolution $P$ of $M$ the space \bea \mathbf{R}\text{Hom}_A(M,N)\eea is defined as the cochain complex \bea 0 \rightarrow \text{Hom}_A(P_0,N) \rightarrow \text{Hom}_A(P_1, N) \rightarrow \dots\eea where the differential is determined from the differential $d$ on $P$. It can be shown that the cohomology of this complex is independent of the chosen resolution $P$, and that the original space $\text{Hom}_A(M,N)$ is recovered by taking the zeroth cohomology.

\paragraph{} The way the differential-graded algebra $A^{!}$ defined above emerges as a space of derived endomorphisms is by considering an explicit projective resolution \bea \text{Bar}(A,A,M) \rightarrow M\eea of $M$ known as the bar complex: \bea \text{Bar}(A,A,M) =  \big\{ \dots \rightarrow A \otimes \bar{A}[-1]^{\otimes 2} \otimes M \rightarrow A\otimes \bar{A}[-1] \otimes M \rightarrow  A \otimes M  \big \} .\eea The differential on this chain complex is given as follows: an element \bea  a\otimes[a_1 | \dots |a_n] \otimes m \in A\otimes \bar{A}[-1]^{\otimes n} \otimes M \eea where we've abbreviated $[a_1 | \dots | a_n] := a_1 \otimes \dots \otimes a_n \in \bar{A}[-1]^{\otimes n}$, is mapped to \bea \begin{split} a a_1\otimes[a_2 | \dots |a_n] \otimes m + \sum_{i=1}^{n-1} (-1)^{i}[a_1| \dots | a_i a_{i+1} | \dots |a_n] \otimes m \\  +  (-1)^n a\otimes[a_1 | \dots |a_{n-1}] \otimes a_{n} \cdot m. \end{split}\eea The map $A\otimes M \rightarrow M$ given by the $A$-module structure on $M$ gives a projective resolution of $M$. The bar resolution of the $A$-module $\mathbb{C}$ gives that $\mathbf{R}\text{Hom}_A(\mathbb{C}, \mathbb{C})$ is the complex \bea \mathbf{R}\text{Hom}_{A}(\mathbb{C}, \mathbb{C}) = \{0 \rightarrow \text{Hom}_A(A \otimes \bar{A}, \mathbb{C}) \rightarrow \text{Hom}_A( A\otimes \bar{A}^{\otimes 2}, \mathbb{C}) \rightarrow \dots \}\eea with differential induced from the one on the bar complex. Since \bea \text{Hom}_A(A\otimes \bar{A}^{\otimes n}, \mathbb{C}) = \text{Hom}_{\mathbb{C}}(A^{\otimes n}, \mathbb{C}) = (\bar{A}^{\vee}[1])^{\otimes n},\eea we get \bea \mathbf{R}\text{Hom}_A(\mathbb{C}, \mathbb{C}) = \{0 \rightarrow \mathbb{C} \rightarrow \bar{A}^{\vee}[1] \rightarrow (\bar{A}^{\vee}[1])^{\otimes 2} \rightarrow \}.\eea The differential on the bar complex gives rise to the differential $Q$ we wrote previously. 

\paragraph{} In summary, letting \bea T^{\bullet}(\bar{A}^{\vee}[1]) = \mathbb{C} \oplus \bar{A}^{\vee}[1] \oplus \big(\bar{A}^{\vee}[1] \big)^{\otimes2} \oplus \dots,\eea be the tensor algebra of the dual of the degree-shifted augmentation ideal of $A$, the Koszul dual  $A^{!}$ is the differential graded algebra \bea A^{!} = \mathbf{R}\text{Hom}_A(\mathbb{C}, \mathbb{C}) = \big(T^{\bullet}(\bar{A}^{\vee}[1]), Q , \cdot) ,\eea where $Q$ is defined by using the product on $\bar{A}$, and $\cdot$ denotes the product of words in the tensor algebra. 

\paragraph{} In general the cobar complex is a very big space (albeit with a simple product) and is not the most useful description of $A^{!}$. In practice one computes the cohomology of the cobar complex \bea \text{Ext}^{\bullet}_{A}(\mathbb{C},\mathbb{C}) = H^{\bullet}_Q\big(\mathbf{R}\text{Hom}_A(\mathbb{C}, \mathbb{C}) \big) \eea to get a much smaller space. The homotopy transfer theorem guarantees that the cohomology $\text{Ext}^{\bullet}_A(\mathbb{C}, \mathbb{C})$ can be equipped with an $A_{\infty}$-structure which is equivalent to the original differential graded algebra $\mathbf{R}\text{Hom}_A(\mathbb{C}, \mathbb{C})$ as an $A_{\infty}$-algebra.

\paragraph{} More generally, given a left $A$-module $M$, we can define a right module for the Koszul dual as \bea \mathbf{R}\text{Hom}_A(M, \mathbb{C}) = \big(M^{\vee} \otimes T^{\bullet}(\bar{A}^{\vee}[1]), Q \big) ,\eea where $Q$ is induced from the algebra module and the algebra multiplication structures. Finally, for two left $A$-modules $M$ and $N$ we can define \bea \mathbf{R}\text{Hom}_A(M, N) =  \big(M^{\vee} \otimes T^{\bullet}(\bar{A}^{\vee}[1]) \otimes N, Q \big)\eea where $Q$ is again determined appropriately from the module structures of $M$ and $N$.

\subsection{Exterior Algebra in Two Variables}

\paragraph{} The cobar complex can be unpacked to compute the Koszul duals of some familiar algebras. 

\paragraph{} Let us warm up by considering a particularly easy example. Take $A$ to be the exterior algebra in one variable $\theta$ \bea A = \mathbb{C}[\theta]/ \langle \theta^2 \rangle\eea so that the augmentation ideal $\bar{A}$ is a one-dimensional space spanned by $\theta$. We assign $\theta$ ghost number unity. Letting $x$ be the Koszul dual variable to $\theta$, the cobar complex is simply the tensor algebra of $\mathbb{C}$ spanned by $x$, \bea A^{!} = T^{\bullet} \mathbb{C},\eea where the ghost number of $x$ is \bea \text{gh}(x) = - \text{gh}(\theta) + 1 = 0.\eea Since $\bar{A}$ is the one-dimensional algebra with trivial multiplication map, the differential on the cobar complex is trivial, so the cohomology coincides with the full tensor algebra $A^{!} = T^{\bullet} \mathbb{C}$. Being a tensor algebra in one variable the product on $A^{!}$ is simply \bea x^n \cdot x^m = x^{n+m} .\eea Thus we recover the standard statement that the Koszul dual of $A = \mathbb{C}[\theta]/ \langle \theta^2  \rangle $ is the polynomial ring: \bea A^{!} = \mathbb{C}[x]\eea

\paragraph{} The converse of this statement is trickier. Take $A =  \mathbb{C}[x]$ equipped with augmentation $x \to 0$. We assign $x$ vanishing ghost number and in addition a ``spin"-grading of $+1$. The augmentation ideal $\bar{A}$ is large: it is spanned by all monomials $x^n$ with $n > 0$. Denote the dual of $x^n$ as $\theta_n$. The space $A^{!}$ is thus the tensor algebra in infinitely many variables $\theta_1, \theta_2, \dots$ where $\theta_n$ carries ghost number $+1$ and spin $n$. The Poincare series \bea P_{A^{!}}(q,t) := \sum_{p,s} \text{dim}(C^{(p,s)})t^p q^s \eea is easily seen to be \bea P_{A^{!}}(q,t) = \frac{1-q}{1-(t+1)q} = 1+ \sum_{n \geq 1} q^n t(1+t)^{n-1},\eea which at $t=-1$ becomes $1-q$, the character of an exterior algebra in one variable having ghost number and spin both being $1$. Cohomologically, the $\theta_1$ generator is $Q$-closed and gives the non-trivial cohomology class corresponding to the original variable $\theta$. The simplest part of the differential is \bea Q \theta_2 = \theta_1^2,\eea so that $\theta_1$ is now only nilpotent up to $Q$-exact terms. More generally \bea Q \theta_n = \sum_{k=1}^{n-1} \theta_k \theta_{n-k}.  \eea The subcomplexes formed by products of $\theta_{n_i}$ such that $\sum_i n_i =n$ for $n >1$ are all acyclic. This is because one can define a contracting homotopy $\widetilde{Q}$ via \bea  \widetilde{Q}(\theta_i) &=& 0, \\ \widetilde{Q}(\theta_{i_1} \theta_{i_2} \dots \theta_{i_n}) &=& \sum_{j=1}^{n-1} (-1)^j\theta_{i_1} \dots \theta_{i_j + i_{j+1}} \dots \theta_{i_n} ,  \,\,\,\,\,\, n \geq 2, \eea so that \bea \{Q, \widetilde{Q} \} \propto 1.\eea  Thus we find that $A^!$ is quasi-isomorphic to $\mathbb{C}[\theta_1]/ \langle \theta_1^2 \rangle$: Koszul dualizing a second time recovers the original algebra up to quasi-ismorphism. 

\paragraph{} For a more complicated but very instructive example, we will work out the Koszul dual of the algebra $A$ generated by two variables $e_1$ and $e_2$ modulo the relations \bea e_1^2 = e_2^2 = e_1e_2 + e_2 e_1 = 0,\eea namely the exterior algebra in two variables $A = \bigwedge^{\bullet} \mathbb{C}^2$. We again assign both $e_1$ and $e_2$ ghost number $+1$. In addition we will also consider a spin grading taking that also to be $+1$ for both $e_1$ and $e_2$. The algebra is four-dimensional with the augmentation ideal being the three-dimensional subalgebra spanned by $e_1, e_2$ and $e_2 e_1$ which are elements of ghost number $1,1$ and $2$ respectively \bea \bar{A} = \text{Span}\{e_1, e_2, e_2 e_1 \}.\eea The Koszul dual algebra, namely the cobar complex $\mathbf{R}\text{Hom}_A(\mathbb{C}, \mathbb{C})$ is thus the tensor algebra in the Koszul dual variables $\alpha_1, \alpha_2$ and $\alpha_{(2,1)}$ where the ghost number and spin of these variables are \bea (\text{gh},s)(\alpha_i) &=& (0,1) \,\,\,\,\,\,\, \text{ for } i =1,2 \\ (\text{gh},s)(\alpha_{(2,1)}) &=& (-1,2).\eea In contrast to the exterior algebra in a single variable, there is now a non-trivial differential that acts via \bea Q \alpha_1 &=& 0, \\ Q \alpha_2 &=& 0, \\ Q \alpha_{(2,1)} &=& \alpha_2 \alpha_1 - \alpha_1 \alpha_2. \eea Since the differential preserves the spin degree, the cobar cochain complex can be organized as a direct sum of finite-dimensional chain complexes for each spin $s$. We denote the $p$th cohomology at spin $s$ as $\text{Ext}^{(p, s)}_A(\mathbb{C}, \mathbb{C}).$ The basic claim is that \bea \text{Ext}_{\bigwedge \mathbb{C}^2}^{(p , s)}(\mathbb{C}, \mathbb{C}) = \begin{cases}  \mathbb{C}^{s+1} \,\,\,\,  \text{ for } p=0 ,& \\ 0 \,\,\,\,\,\,\,\,\,\,\,\,\,\,\,\, p \neq 0 .\end{cases} \eea Moreover, the $s+1$ cohomology classes are given by the classes of \bea \alpha_1^{s}, \,\, \alpha_1^{s-1}\alpha_2, \,\, \alpha_1^{s-2} \alpha_2^2, \dots, \, \alpha_2^{s}.\eea This is easy to verify directly for low enough spin. The cases $s=0$ and $s=1$ hold trivially, since the differential acts trivially on $1, \alpha_1, \alpha_2$. The first non-trivial cancellation occurs at $s=2$ where we have a complex of the form \bea \begin{CD} 0 @>>> \mathbb{C} @>Q>> \mathbb{C}^4 @>>> 0, \end{CD}\eea being spanned by $\alpha_{(1,2)}$ in ghost number $-1$ and $\alpha_1^2, \alpha_{1}\alpha_2, \alpha_2 \alpha_1, \alpha_2^2$ in ghost number $0$. Since $Q$ acts non-trivially on $\alpha_{(1,2)}$ the cohomology is automatically trivial in ghost number $-1$ and three-dimensional in ghost number $0$, which agrees with the claim above at $s=2$. We can keep pushing and directly look at $s=3$ where now the complex is of the form \bea \begin{CD} 0 @>>>  \mathbb{C}^4  @>Q>> \mathbb{C}^8 @>>> 0 \end{CD} \eea where $Q$ acts surjectively, so we again find the expected answer. As we go to higher spin the complexes become much larger: the Poincare series of $A^{!}$ is given by \bea P_{A^{!}}(q,t) &=& (1-2q - q^2t^{-1})^{-1}, \\ &=& \sum_{s=0}^{\infty} q^s \sum_{j=0}^{\lfloor \frac{s}{2} \rfloor} 2^{s-2j} t^{-j} \binom{s-j}{j}  \eea which determines the dimensions of the various complexes \bea P_{A^{!}}(q,t) = 1 + 2q + (t^{-1} + 4)q^2 + (4t^{-1} + 8)q^3 + (t^{-2} + 12t^{-1} + 16)q^4 + ...\eea At $t=-1$ this becomes \bea P_{A^{!}}(q,-1)=  (1-q)^{-2}, \eea the character of a symmetric algebra in two variables, each carrying vanishing ghost number and unit spin, consistent with the claim above. 

\paragraph{} To prove the above claim holds cohomologically the basic technique is to define a filtration of the cobar complex. A filtration of a chain complex consists of a collection of subcomplexes $\{F_a \}_{a \in I}$ labeled by an ordered set $I$ such that $F_a \subseteq F_b$ if $a<b$. Being a subcomplex means that $F_a$ is an invariant subspace for the differential $d$ for each $a \in I$. In the present case a natural filtration of the spin $s$ complex can be defined as follows. We note that each element in the spin $s$ complex determines a unique label of the form \bea (i_1, \dots, i_s) \eea where each $i_j$ is either $1$ or $2$. For instance, the following word of spin $5$ \bea \alpha_{1} \alpha_{(2,1)} \alpha_2 \alpha_2\eea has a labeling of $$(1,2,1,2,2).$$ By setting $1<2$ we can define an ordering on the set of $s$-tuples by using the lexicographic ordering. The largest element in such an ordering is of course the element with all $2$s. We set $F_a$ for a given $s$-tuple $a = (i_1, \dots, i_s)$ to be the subspace spanned by all words with an $a$-label equal to or less than $a$. For instance at $s=2$, we have $F_{(1,1)}$ is the one-dimensional space spanned by $\alpha_1^2$, $F_{(1,2)}$ adjoins to it the element $\alpha_{1} \alpha_2$, $F_{(2,1)}$ further adjoins two additional elements $\alpha_2 \alpha_2$ and $ \alpha_{(2,1)}$ and $F_{(2,2)}$ finally adjoints to it $\alpha_{2} \alpha_2$ to obtain the full filtration \bea F_{(1,1)} \subset F_{(1,2)} \subset F_{(2,1)} \subset F_{(2,2)} = C^{(\bullet, s=2)} \eea of the spin $2$ complex.  Importantly, since $Q\alpha_{(2,1)} = \alpha_{2} \alpha_1 - \alpha_1 \alpha_2$ these subspaces are subcomplexes. 

\paragraph{} A filtered complex determines a spectral sequence which converges to the associated graded space of the induced filtration on the cohomology. The associated graded space of a filtered space is simply the space \bea G C = \bigoplus_{a \in I} F_a/F_{a-1} =: \oplus_{a \in I} V_a\eea where \bea F_{a-1} = \oplus_{b < a} F_{b}.\eea The differential induces a differential $[d_a]$ on $V_a$, the $a$th graded component of the space $GC$. The first non-trivial approximation (formally the first page of the spectral sequence of a filtered complex) simply considers the cohomology of $V_a$ with respect to the induced differentials \bea \mathcal{E}_1^{(a, \bullet)} = H^\bullet(V_a, [d_a]). \eea

\paragraph{} In the present case the filtering induced by the lexicographic ordering leads to a particularly nice situation. A given $s$-tuple $(i_1, \dots, i_s)$ determines a unique integer $k$ between $1$ and $\lfloor \frac{s}{2} \rfloor$ defined to be the number of times a $2$ is followed by a $1$ right after. For instance the $7$-tuple \bea (1,2,2,1,1, 2,1) \eea has $k=2$.  We claim that the associated graded chain complex has $V_{(i_0, \dots, i_n)}$ has a simple description. As a vector space it is spanned by elements carrying precisely the tuples labeled as above. Such a word is formed by making a choice at each appearance of a $(2,1)$: we choose to put either an $\alpha_2 \alpha_1$ or an $\alpha_{(2,1)}$. Everything else is uniquely fixed by the $I$-label. For the same $7$-tuple as above for instance there are a total of four words given by \bea \alpha_1 \alpha_2 \alpha_2 (\alpha_2 \alpha_1) \alpha_1 (\alpha_2 \alpha_1), \,\,\,\,  \alpha_1 \alpha_2 \alpha_2 (\alpha_{(2,1)}) \alpha_1 (\alpha_2 \alpha_1), \\  \alpha_1 \alpha_2 \alpha_2 (\alpha_2 \alpha_1) \alpha_1 (\alpha_{(2,1)}), \,\,\,\,\,  \alpha_1 \alpha_2 \alpha_2 (\alpha_{(2,1)}) \alpha_1 (\alpha_{(2,1)}),\eea where the ghost numbers are \bea 0, -1,-1, -2\eea respectively. For a generic tuple $I$ with $k(I) = k$, we get a $2^k$-dimensional cochain complex of the following form \bea 0 \rightarrow \mathbb{C} \rightarrow \mathbb{C}^k \rightarrow \mathbb{C}^{\binom{k}{2}} \rightarrow \dots \rightarrow \mathbb{C}^k \rightarrow \mathbb{C} \rightarrow 0,\eea with Poincare polynomial \bea P_{V_{I}}(t) = (t^{-1}+1)^k. \eea Letting $x = \alpha_{(2,1)}$ and $y=\alpha_2 \alpha_1$ it can be described as the space of length $k$ words in the non-commuting variables $x$ and $y$ (with ghost numbers $-1$ and $0$ respectively) with the differential being determined by \bea Qx &=& y, \\ Qy &=& 0,\eea and extending by the Leibniz rule. This cochain complex is acylic because the degree $-1$ operator defined via \bea \widetilde{Q} x &=& 0, \\ \widetilde{Q} y &=& x\eea and extended to act via the Leibniz rule provides a contracting homotopy \bea \{Q, \widetilde{Q} \} \propto 1.\eea Thus every graded complex $V_I$ associated to an $s$-tuple $I$ is acylic provided $k(I) \geq 1$. If $k=0$ then the word is of the form \bea (1, \dots, 1, 2, \dots,2) \eea and the associated graded complex is one-dimensional spanned by a class of the form \bea \alpha_1 \dots \alpha_1 \alpha_2 \dots \alpha_2.  \eea There are precisely $s+1$ such classes. The spectral sequence of the filtration collapses at the first page and this therefore proves the claim.

\paragraph{} Finally the product of two classes can be determined: we have the product of polynomials up to $Q$-exact terms. The algebra is formal because the cohomology is concentrated in a single degree and there can be no higher products. This shows that the Koszul dual is the polynomial ring in two variables \bea A^{!} \simeq \mathbb{C}[\alpha_1, \alpha_2].\eea

\subsection{Koszul Duals of PBW Algebras}

\paragraph{} The reason we went through the above calculation in such detail is that the reasoning used to derive the Koszul dual generalizes to a much wider class of algebras. These are algebras with a so-called Poincare-Birkhoff-Witt (PBW) basis. An algebra \bea A = T^{\bullet}V/ \langle S \subset V\otimes V \rangle\eea defined by an ordered set of generators $\{e_1,e_2, \dots \}$ (equivalently a basis of $V$) and \textit{quadratic} relations defined by a subspace of $V\otimes V$ spanned by elements of the form \bea v_{kl} := f^{ij}_{kl} \, e_{i}\otimes e_{j} \eea is said to be a \textit{PBW algebra} if it has a basis \bea \{e_{i_1} \dots e_{i_n} \}\eea such that the labels $(i_1, \dots, i_n)$ occuring in the basis obey the following properties: first we require that the product of two basis elements $(i_1, \dots, i_n)$ and $(j_1, \dots, j_m)$ is either another basis element, or can be expressed as words which carry labels strictly greater than $(i_1, \dots, i_2, j_1, \dots, j_m)$ in the lexicographic ordering. Secondly, we require that if $(i_1, \dots, i_n)$ is a basis element then so is $(i_1, \dots, i_k)$ and $(i_{k+1}, \dots, i_n)$ for each $1\leq k \leq n$. 

\paragraph{} In particular for two generators $e_i$ and $e_j$ we must have that either $e_i e_j$ is a basis element, or \bea e_i e_j = \sum_{(r,s) > (i,j)} c_{ij}^{rs}e_r e_s \eea for some constants $c_{ij}^{rs}$. Suppose $B_2$ is the set of labels that occurs in a PBW basis of quadratic elements \bea B_2 =\{ (a,b)| e_a e_b \text{ is a basis element } \} .\eea The subspace $S$ is thus generated by elements of the form \bea v_{ij} = e_{i}e_{j} - \sum_{(r,s)>(i,j)} c_{ij}^{rs} e_{r}e_{s},\eea where $(i,j)$ is not an element of $B_2$, i.e $e_i e_j$ does not occur as a labeled basis element. 

\paragraph{} For instance in the exterior algebra in $n$ variables, we can take a basis of quadratic elements \bea \{e_i e_j \} \eea where $i > j$ so that \bea B_2= \{(i,j) | i> j, i,j \in \{1, \dots, n\} \} .\eea Then the labels that don't occur are elements $(i,j)$ with $i \leq j$. These impose the relations which impose the relations \bea e_{i}e_j = -e_j e_i\eea giving us a $\frac{n(n+1)}{2}$-dimensional subspace of relations. 

\paragraph{} Given a PBW algebra Priddy \cite{priddykoszul} showed the following fundamental result: the Koszul dual of a PBW algebra \bea A = T^{\bullet} V/ \langle S \rangle\eea where $S \subset V \otimes V$ is the subspace of relations, is given by \bea A^{!} \simeq A^![e] \equiv T^{\bullet} V^{\vee}[1]/ \langle \text{Ann}(S) \rangle\eea where $\text{Ann}(S)\subset V^{\vee}[1] \otimes V^{\vee}[1]$ is the subspace which annihilates $S$. In other words, for PBW algebras, the quadratic dual coincides with the Koszul dual up to quasi-isomorphism. We introduced the notation $A^![e]$ to denote the result of dualizing $A$ in a specific PBW basis $e$. If $A$ admits two PBW presentations
with bases $e$ and $e'$, we automatically get a quasi-isomorphism $A^![e] \simeq A^![e']$.  

\paragraph{} The way one proves this is very similar to the computation of the Koszul dual of exterior algebra in two variables. The presence of a PBW basis allows one to filter the cobar complex by lexicographic ordering of the labeled basis. One shows that the only non-trivial cohomology classes are generated by variables $\alpha^i$ dual to the $e_i$ \, \footnote{A quadratic algebra such that the cohomology of the cobar complex is generated by the dual generators is called homogeneous Koszul. Another way of stating Priddy's result is to say that PBW algebras are homogeneous Koszul.}. The product of such classes is determined by the relations imposed by \bea Q \alpha^{(i,j)} = 0,\eea where $\alpha^{(i,j)}$ is the dual variable to $e_i e_j$. Explicitly, there's a relation for each $(i,j) \in B_2$. Since we have \bea Q \alpha^{(i,j)} = \alpha^i \alpha^j + \sum_{(r,s)<(i,j)} c_{rs}^{ij} \alpha^r \alpha^s\eea the dual space of relations for the quadratic dual is spanned by \bea v^{ij} = \alpha^i \alpha^j + \sum_{(r,s)<(i,j)} c_{rs}^{ij} \alpha^r \alpha^s .\eea This precisely coincides with $\text{Ann}(S) \subset (V^{\vee}[1])^{\otimes 2} $. 
  
\paragraph{} In what follows we will be needing a slight generalization of this result, which was also discussed in \cite{priddykoszul}. This is the following. We suppose that we have an algebra $A$ with generators $\{e_1, e_2, \dots \}$ subject to certain relations: \bea A = T^{\bullet}V /S \eea where now $$S \subset V\oplus V^{\otimes 2}.$$ In other words, we now allow linear terms in the defining relations for $A$. $A$ has a filtration where $F_k$ is spanned by words in the tensor algebra of length less than or equal to $k$. Let $GA$ be the associated graded algebra. We will suppose that $GA$ is a PBW algebra. More explicitly, this means that if as before $B_2$ is the set of basis labels for quadratic elements $e_i e_j$ of $A$ (more precisely, a basis of $G_2 = F_2/F_1$), then we have relations in $A$ of the form  \bea e_i e_j = f_{ij}^k e_k + \sum_{(r,s)>(i,j)} c_{ij}^{rs} e_r e_s  \eea for each $(i,j) \notin B_2$. A ``quadratic dual'' dg-algebra $A^![e]$ quasi-isomorphic to $A^!$ can now be explicitly described as a differential graded algebra generated by the Koszul dual variables $\{\alpha^i\}$ subject to the differential \bea Q \alpha^i = \sum_{(j,k) \notin B_2}f^i_{jk} \alpha^j \alpha^k\eea and the relations \bea  \alpha^i \alpha^j + \sum_{(r,s)<(i,j)} c_{rs}^{ij} \alpha^r \alpha^s = 0. \eea This generalization to include linear terms allows us to deduce the well-known Koszul dual pair associated to a Lie algebra $\mathfrak{g}$: the Koszul dual of the universal enveloping algebra $U(\mathfrak{g})$ is the Chevalley-Eilenberg differential graded algebra $\bigwedge \mathfrak{g}^{\vee}[1]$ \bea \big(U(\mathfrak{g}) \big)^! \simeq \Big(\bigwedge \mathfrak{g}^{\vee}[1], \, Q_{\text{CE}} \Big).\eea The linear terms in the relation \bea e_a e_b - e_b e_a = f_{ab}^c e_c\eea imposed by the Lie bracket Koszul dualize to the Chevalley-Eilenberg differential \bea Q_{\text{CE}} \, \alpha^a = \frac{1}{2} f^{a}_{bc} \, \alpha^b \alpha^c. \eea \paragraph{} In the next section we will apply these formulas to deduce the Koszul dual of $\mathcal{H}(A_2)$. 

\paragraph{}The quadratic $\oplus$ linear case will be sufficient for the purposes of this paper. It may be that a more general choice of a PBW basis $A_\vartheta$ 
requires cubic and higher terms in the $ee$ exchange relations. If so, $A^!_\vartheta[e]$ will be an $A_\infty$ algebra quasi-isomorphic to $(A_{\vartheta})^!$.
\subsection{On Characters}
In our application, $A=A_\vartheta$ is graded by $\Gamma$ and $\bZ$, in the sense that the algebra operation is a degree preserving map $A \, \widehat \otimes A \rightarrow A$. The equivariant character of $A_\vartheta$ satisfies 
\begin{equation}
\chi_{A_\vartheta}=\mathcal{S}^{-1}_\vartheta \,.
\end{equation}
so that $A_\vartheta$ categorifies the inverse quantum spectrum generator. 
It is clear from the definition that 
\begin{equation}
\chi_{A^!_\vartheta}=\mathcal{S}_\vartheta \,.
\end{equation}
so that $A^!_\vartheta$ or dg-algebras quasi-isomorphic to it are possible categorifications of the quantum spectrum generator. 

\paragraph{} We expect (and see concretely in examples) a factorization
\begin{equation}
A_\vartheta = \bigotimes^{\cwarrow}_{\vartheta+\pi > \arg Z_\gamma \geq \vartheta} A_\gamma
\end{equation}
i.e. to have a collection of sub-algebras $A_\gamma$ and a linear basis for $A_\vartheta$ consisting of products of basis elements in $A_\gamma$'s, ordered according to the phase of the central charges. 

\paragraph{} We expect  
\begin{equation}
\chi_{A_\gamma}=\mathcal{K}^{-1}_\gamma \,.
\end{equation}
so that this presentation of $A_\vartheta$ categorifies the factorization of $\mathcal{S}^{-1}_\vartheta$.
The fermionic wall-crossing formula is categorified by the algebra isomorphism between different presentations 
of $A_\vartheta$.

\paragraph{} We do not know a priori if the $A_\gamma$ themselves would be PBW algebras, but we will temporarily assume so. This will be the case in our example below. Then the dual algebra $A^!_\vartheta[e]$ built from the PBW basis will also factor as 
\begin{equation}
A^!_\vartheta[e] = \bigotimes^{\cwarrow}_{\vartheta \leq \arg Z_\gamma < \vartheta+\pi} A^!_\gamma[e]
\end{equation}
This factorization categorifies the factorization of $\mathcal{S}_\vartheta$! The bosonic wall-crossing formula is thus categorified by the quasi-isomorphism between different $A^!_\vartheta[e]$'s.

\paragraph{} We expect the PBW approach to Koszul duality to hold also in situations where the individual $A_\gamma$'s are not PBW algebras, so that a quasi-isomorphic version of $A^!_\vartheta$ may still be expressed as an ordered product of 
$A^!_\gamma$'s.

\pagebreak

\section{The Koszul Dual of $\mathcal{H}(A_2)$} \label{sec:dualize} Having discussed the aspects of Koszul duality that we need it is now a straightforward exercise to work out the Koszul duals of the cohomological Hall algebra for the $A_2$ quiver $\mathcal{H}(A_2)$. The description of the Koszul dual algebra we will recover will depend of course on the presentation. Throughout the section it will be convenient for us to denote the Koszul dual $\alpha_n$ of a generator of $e_n$ by \bea \alpha_n = \frac{1}{n!} (-1)^n \del^n \alpha.\eea

\subsection{Two-Particle Chamber} We begin with recalling the presentation of the COHA in the two particle chamber. Here we have an algebra with generators \bea \{e^1_0, e^1_1, e^1_2, \dots, e^2_0, e^2_1, e^2_2, \dots \}\eea subject to the COHA relation \eqref{cohaexchange}. The basis of elements \bea e^2_{n_1} \dots e^2_{n_k} e^1_{m_1} \dots e^2_{m_l} \eea where $n_1 > \dots > n_k$ and $m_1> \dots > m_k$ forms a PBW basis: if we order $\gamma_2 > \gamma_1$, so that there is a basis of quadratic elements of the form \bea \{e^2_n e^2_m, e^1_n e^1_m, e^2_n e^1_m \}.\eea The relations \bea e^1_n e^2_m = e^2_{m+1} e^1_n - e^2_m e^1_{n+1}\eea guarantee the PBW property.

\paragraph{} The Koszul dual algebra is therefore completely straightforward to describe. It will have generators \bea e^1_n \leftrightarrow \del^n a_{\gamma_1}, \\ e^2_{n} \leftrightarrow \del^n a_{\gamma_2}.\eea Since the entire COHA is concentrated in ghost number zero, the generators $\del^n a_{\gamma_1}$ and $\del^n a_{\gamma_2}$ carry ghost number $+1$. The spins are the same: both $\del^n a_{\gamma_1}$ and $\del^n a_{\gamma_2}$ carry spin $n + \frac{1}{2}$. These generators are subject to the relations \bea \del^n a_{\gamma_1} \del^m a_{\gamma_1} &=& \del^m a_{\gamma_1} \del^n a_{\gamma_1}, \\ \del^n a_{\gamma_2} \del^m a_{\gamma_2} &=& \del^m a_{\gamma_2} \del^n a_{\gamma_2}, \eea which simply states that $\del^n a_{\gamma_1}$ and $\del^m a_{\gamma_2}$ individually form an infinite polynomial algebra, along with the dual of the relation \eqref{cohaexchange} which reads \bea \frac{1}{n!}\del^n a_{\gamma_2} \frac{1}{m!}\del^m a_{\gamma_1} =  \frac{1}{m!} \del^m a_{\gamma_1} \frac{1}{(n-1)!} \del^{n-1} a_{\gamma_2}-\frac{1}{(m-1)!} \del^{(m-1)} a_{\gamma_1} \frac{1}{n!} \del^{n} a_{\gamma_2} , \eea which is simply \bea  \del^n a_{\gamma_2}\del^m a_{\gamma_1} = n \del^m a_{\gamma_1}  \del^{n-1} a_{\gamma_2} -m \del^{m-1} a_{\gamma_1}  \del^{n} a_{\gamma_2} .\eea These relations take a particularly nice form in terms of the generating functions: In addition to the standard bosonic relations \bea a_{\gamma_1}(z) a_{\gamma_1}(w) = a_{\gamma_1}(w) a_{\gamma_1}(z),\\ a_{\gamma_2}(z) a_{\gamma_2}(w) = a_{\gamma_2}(w) a_{\gamma_2}(z), \eea the exchange relation takes the form \bea \label{exch1} a_{\gamma_2}(z) a_{\gamma_1}(w) = (z-w) a_{\gamma_1}(w) a_{\gamma_1}(z).\eea The exchange relation tells us that we have a vector space built from generating functions of the form \bea a_{\gamma_1}(z_1) \dots a_{\gamma_1}(z_n) a_{\gamma_2}(w_1) \dots a_{\gamma_2}(w_m)\eea with the correct gauge charges, spin and ghost numbers. Thus as a vector space we have recovered precisely the space $$\mathcal{C}_{\text{L}} =\mathcal{C}_{\gamma_1} \widehat{\otimes} \,\mathcal{C}_{\gamma_2}.$$ In addition Koszul duality has allowed us to define a natural associative product on this space via the relation \eqref{exch1}, turning it into a graded associative algebra.

\subsection{Three-Particle Chamber} We shall now Koszul dualize the alternate presentation of $\mathcal{H}(A_2)$ which we recall consisted of three sets of fermionic generators \bea \{e^{1}_0, e^1_1, e^1_2, \dots, e^{(12)}_0, e^{(12)}_1, e^{(12)}_2,\dots, e^2_0, e^2_1, e^2_2, \dots\}\eea subject to the relations \eqref{rel1}, \eqref{rel2}, \eqref{rel3}. By choosing the ordering \bea \gamma_1 < \gamma_1 + \gamma_2 < \gamma_2\eea it is straightforward to see that the elements \bea \{e^{1}_{p_1} \dots e^1_{p_k} e^{(12)}_{q_1} \dots e^{(12)}_{q_l} e^{2}_{r_1} \dots e^2_{r_m} \}\eea form a PBW basis for the associated graded algebra, and we can thus apply the results of \cite{priddykoszul}, summarized above, to compute the Koszul dual.

\paragraph{} The Koszul dual algebra to this presentation of the COHA has generators \bea \{\del^n \alpha_{\gamma_1}, \del^n \alpha_{\gamma_1 + \gamma_2}, \del^n \alpha_{\gamma_2} \}_{n=0,1,2,\dots}\eea which are the dual variables to \bea \{e^1_n, e^{(12)}_n, e^2_n\}_{n=0,1,2,\dots} \eea respectively. Since the three sets of $e$'s individually obey fermionic statistics, the three sets of $\alpha$'s individually obey bosonic statistics \bea \del^n \alpha_{\gamma} \del^m \alpha_{\gamma} = \del^m \alpha_{\gamma} \del^n \alpha_{\gamma} \,\,\, \text{ for } \,\,\, \gamma \in \{ \gamma_1, \gamma_1 + \gamma_2, \gamma_2 \}.\eea In addition, we must dualize the exchange relations \eqref{rel1}, \eqref{rel2}, \eqref{rel3}. The dual exchange relations read \bea \frac{1}{n!} \del^n \alpha_{\gamma_1} \frac{1}{m!} \del^m \alpha_{\gamma_2} &=& \sum_{k=0}^n \frac{1}{(n-k)! (m+k+1)!} \del^{(m+k+1)} \alpha_{\gamma_2} \del^{(n-k)} \alpha_{\gamma_1}, \\ \frac{1}{n!} \del^n \alpha_{\gamma_1+ \gamma_2}\frac{1}{m!} \del^m \alpha_{\gamma_2} &=& -\sum_{k=0}^n \frac{1}{(n-k)! (m+k+1)!} \del^{(m+k+1)} \alpha_{\gamma_2} \del^{(n-k)} \alpha_{\gamma_1 + \gamma_2}, \\ \frac{1}{n!} \del^n  \alpha_{\gamma_1} \frac{1}{m!} \del^m \alpha_{\gamma_1 + \gamma_2} &=&  \begin{split} \sum_{k=0}^n \frac{1}{(n-k)! (m+k+1)!} \Big(-\del^{(m+k+1)} \alpha_{\gamma_1+\gamma_2} \del^{(n-k)} \alpha_{\gamma_1}\,\,\,\,\,\,\,\,\,\,\, \\ + \del^{(n-k)}\alpha_{\gamma_1+ \gamma_2} \del^{(m+k+1)} \alpha_{\gamma_1} \Big). \end{split}\eea Finally, the linear term in the relation \eqref{rel1} gives rise to a non-trivial differential in the Koszul dual that reads \bea Q \left(\frac{1}{n!} \del^n\alpha_{\gamma_1+ \gamma_2} \right) = \sum_{k=0}^n \frac{1}{k!} \frac{1}{(n-k)!}\del^{k} \alpha_{\gamma_2} \del^{n-k} \alpha_{\gamma_1},\eea which is simply \bea Q \big( \del^n\alpha_{\gamma_1+ \gamma_2} \big) = \sum_{k=0}^n \binom{n}{k}\del^{k} \alpha_{\gamma_2} \del^{n-k} \alpha_{\gamma_1}. \eea Thus we see that the linear term in \eqref{rel1} encouragingly reproduces \eqref{diffonder}. 

\paragraph{} The relations above can be nicely recast in terms of generating functions. There they read \bea \label{dualrel1} \alpha_{\gamma_1}(z) \alpha_{\gamma_2}(w) &=& \frac{1}{w-z}\big(\alpha_{\gamma_2}(w) - \alpha_{\gamma_2}(z) \big) \alpha_{\gamma_1}(z), \\ \label{dualrel2} \alpha_{\gamma_1+ \gamma_2}(z) \alpha_{\gamma_2}(w) &=& \frac{1}{z-w} \big(\alpha_{\gamma_2}(w) - \alpha_{\gamma_2}(z) \big) \alpha_{\gamma_1 +\gamma_2}(z), \\ \label{dualrel3} \alpha_{\gamma_1}(z) \alpha_{\gamma_1+ \gamma_2}(w) &=& \frac{1}{z-w} \big( \alpha_{\gamma_1+\gamma_2}(w) \alpha_{\gamma_1}(z) - \alpha_{\gamma_1+ \gamma_2}(z) \alpha_{\gamma_1}(w) \big). \eea We note that the numerators of the expressions that appear on the hand sides vanish when $z=w$ so the singularities are spurious as they should be. In addition the differential in terms of the generating series is \bea Q\big(\alpha_{\gamma_1+ \gamma_2}(z) \big) = \alpha_{\gamma_2}(z) \alpha_{\gamma_1}(z).\eea

\paragraph{} Koszul duality guarantees that these relations along with the differential define a consistent differential graded algebra. As a check on our formulas one can verify explicitly that associativity holds: for instance we have \bea \big(\alpha_{\gamma_1}(z) \alpha_{\gamma_1+\gamma_2}(w) \big) \alpha_{\gamma_2}(x) = \alpha_{\gamma_1}(z) \big(\alpha_{\gamma_1+\gamma_2}(w) \alpha_{\gamma_2}(x) \big).   \eea In addition we can also formulate the differential on an arbitrary state and check its nilpotence. 

\paragraph{Explicit Formula for $Q$:} Let us work out the action of the differential on an arbitrary state. It is not difficult to show that a repeated application of the Leibniz rule along with the relations \eqref{dualrel1}, \eqref{dualrel3} give us a simple explicit formula for the differential: \bea \label{qkoszul} Q_{\text{K}} \big(\alpha_{\gamma_1+ \gamma_2}(z_1) \dots \alpha_{\gamma_1+\gamma_2}(z_n) \big) = \sum_{i=1}^n \alpha_{\gamma_2}(z_i) \widehat{\otimes}\,\prod_{j\neq i} \frac{\alpha_{\gamma_1+ \gamma_2}(z_j)}{z_i -z_j}  \widehat{\otimes}\,\alpha_{\gamma_1}(z_i) .\eea We denote this differential as $Q_{\text{K}}$ (where K stands for Koszul) in order to distinguish it from the differentials that have appeared above.  It is easy to check that $Q_{\text{K}}$ is nilpotent. Once again all singularities in the formula are spurious.

\paragraph{} We observe that the differential $Q_{\text{K}}$ above takes the same general form as the one we wrote previously \eqref{pentagondifferential1} with the replacement \bea (\del_i - \del_j) \rightarrow (z_i-z_j)^{-1}.\eea More precisely we claim that there's a chain map between the complex with the differential in \eqref{pentagondifferential1} and the complex with differential \eqref{qkoszul}. The explicit chain map acts as the identity on the $\alpha_{\gamma_1}$ and $\alpha_{\gamma_2}$, and acts on products of the $\alpha_{\gamma_1+\gamma_2}$ states by \bea \phi \big(\alpha_{\gamma_1+\gamma_2}(z_1)\dots \alpha_{\gamma_1+\gamma_2}(z_n) \big) = \prod_{i \neq j} \big[(\del_i - \del_j) \big] \big( (z_i -z_j) \alpha_{\gamma_1+\gamma_2}(z_1) \dots \alpha_{\gamma_1 + \gamma_2}(z_n) \big). \eea Note that there is no sign ambiguity in the formula, and that it indeed preserves all degrees. It is easy to verify that \bea Q_{\text{K}} \circ \phi = \phi \circ Q_{\text{R}} \eea namely it is a chain map between the two complexes. 

\paragraph{} Since Koszul dual algebras of isomorphic algebras are quasi-isomorphic, this proves the categorified bosonic pentagon identity.

\pagebreak

\section{Web Interpretation} \label{sec:web} We now interpret our main formulas; the product in the two-particle chamber, and the differential-graded algebra structure in the three-particle chamber, in terms of a web-like formalism. We will only initiate the discussion here, leaving the general development of these ideas to future work.

\paragraph{} We begin by recalling that the BPS particles of two-dimensional theories with complex central charges can be nicely organized to build bulk and boundary algebras of observables. This is done via a so-called ``web formalism," developed in \cite{Gaiotto:2015aoa}. Central to the formalism are a collection of Feynman-like diagrams involving BPS particles propagating at angles which correspond to the arguments of their central charge. One recovers boundary $A_{\infty}$-categories by considering web diagrams on a half-plane and similarly bulk $L_{\infty}$-algebras by considering webs on the plane. 

\paragraph{} Four-dimensional $\mathcal{N}=2$ theories have a few aspects in common with the simpler case of two-dimensions, with one of the main similarities being that they also have BPS particles with complex central charges. One can pursue the analogy further by considering the holomorphic-topological twist of the four-dimensional theory, since then there is a sensible splitting of the four directions into two real topological directions and one complex holomorphic direction \bea \mathbb{R}^4 = \mathbb{R}^2 \times \mathbb{C}.\eea We can also similarly consider boundaries taken to be supported along one real topological direction and the one complex direction. By projecting the trajectories of BPS particles to the topological plane and requiring that the projection of the trajectory of a particle with charge $\gamma$ be aligned with the central charge of $Z_{\gamma}$ one can sensibly begin formulating BPS webs in the present context.

\begin{figure}
\centering
\includegraphics[width=0.75\textwidth]{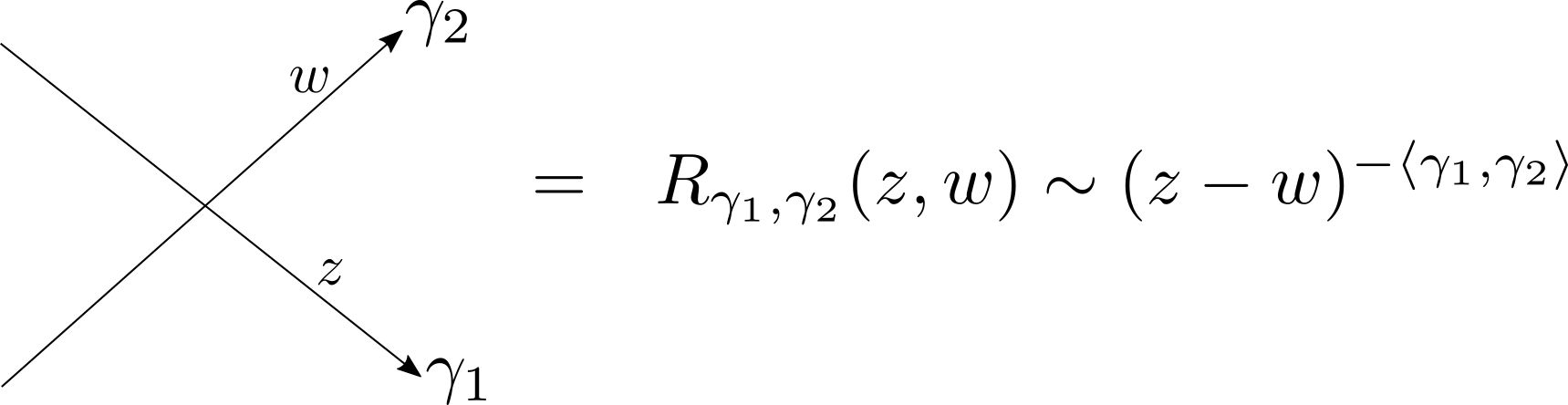}
\caption{A proposed rule for the crossing of two BPS particles of charges $\gamma_1$ and $\gamma_2$. }
\label{rmatrix}
\end{figure}

\paragraph{} Naturally one can anticipate a few differences from the simpler, two-dimensional formalism. First, in the holomorphic-topological setting each particle will carry a label $z \in \mathbb{C}$ which indicates its position in the holomorphic plane, which will be taken to be fixed. The main difference from two dimensions is that there can now be crossing phenomenon\footnote{Not to be confused with \textit{wall}-crossing phenomenon, which occurs in both two and four dimensions.}: the trajectories of two BPS particles carrying labels $z$ and $w$ is said to cross if the projection of their trajectories intersects in the topological plane.   Second, the IR theory is not gapped, but rather is an Abelian gauge theory. The HT twist of the Abelian gauge theory is a $BF$ version of 4d HT Chern-Simons theory and thus massive dyons will behave as 't Hooft-Wilson lines in the $BF$ theory. 

\paragraph{} The crossing of two BPS particles of charge $\gamma$ and $\gamma'$ carrying spectral labels $z$ and $w$ respectively can be imagined as the crossing of two Abelian Wilson-`t Hooft lines in the Abelian holomorphic-topological BF theory. This is analogous to the crossing of a Wilson and `t Hooft line in four-dimensional Chern-Simons theory \cite{Costello:2021zcl}. We claim that in the present case the natural R-matrix that controls such crossings is \bea R_{\gamma, \gamma'}(z,w) = (z-w)^{-\langle \gamma, \gamma' \rangle},\eea and that this R-matrix should appear in the amplitude for any web diagram that involves such a crossing. We anticipate that this R-matrix is a fundamental new ingredient in going from the two-dimensional topological web formalism to its four-dimensional holomorphic-topological cousin. Another ingredient is the inclusion of photons in the operator algebra \cite{future}. We disregard that (important) contribution in this paper.  

\begin{figure}
\centering
\includegraphics[width=0.73\textwidth]{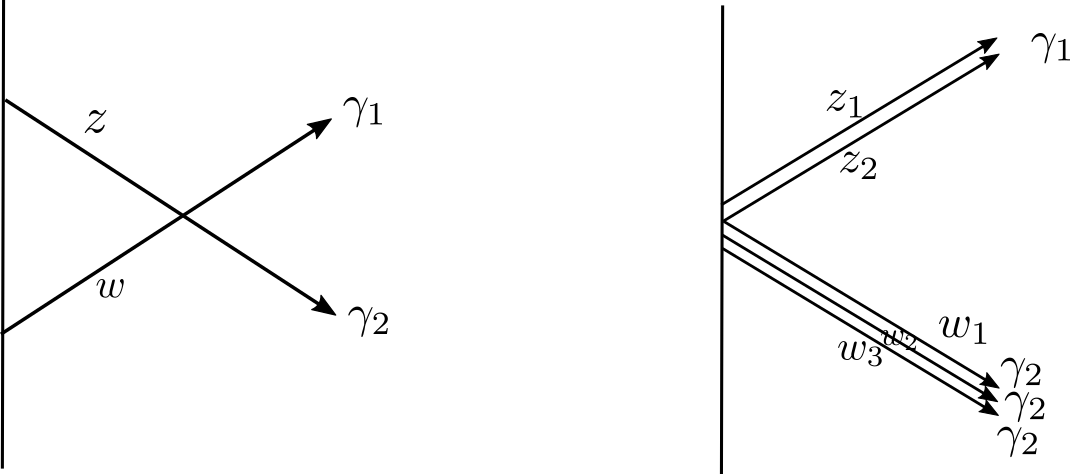}
\caption{The left depicts a crossing diagram leading to the multiplication rule given by \eqref{2particlerel}. The right depicts a BPS half-fan corresponding to states of gauge charge $2\gamma_1 + 3\gamma_2$ namely states of the form $a_{\gamma_1}(z_1) a_{\gamma_1}(z_2) a_{\gamma_2}(w_1) a_{\gamma_2}(w_2) a_{\gamma_2}(w_3)$. }
\label{2particlediagrams}
\end{figure}

\paragraph{} This crossing R-matrix appears sufficient to give a web description of the algebra on the two-particle side of the categorified bosonic pentagon identity. Identify $a_{\gamma_i}(z)$ as a local operator which creates a BPS particle 
of charge $\gamma_i$ at $z$. When we multiply $a_{\gamma_1}(w)$ and $a_{\gamma_2}(z)$, the projected trajectories of the resulting particles will either cross or not, depending on the order of the two factors. This reproduces the crucial relation 
\bea \label{2particlerel} a_{\gamma_2}(z) a_{\gamma_1}(w) = (z-w) a_{\gamma_1}(w) a_{\gamma_2}(z) \eea defining the algebra, as \bea R_{\gamma_2, \gamma_1}(z,w) = (z-w)^{-\langle \gamma_2, \gamma_1 \rangle} = (z-w) \, .\eea Either side is identified with the boundary local operator associated to a ``half fan'' of two BPS particles. See Figure \ref{2particlediagrams}.

\paragraph{} A generic element of the algebra can be obtained by expanding out \bea \prod_i a_{\gamma_2}(z_i) \prod_j a_{\gamma_1}(w_j), \eea which can be identified as the boundary local operator associated to a generic ``half fan'' of BPS particles. 

\paragraph{Remark} The R-matrix coming from the crossing amplitude of BPS particles and how it gives rise to a non-trivial product in the above example is somewhat reminiscent of Harvey-Moore BPS algebras \cite{Harvey:1995fq, Harvey:1996gc, Moore:1997ar}. It would be interesting to relate in a more general setting, cohomological Hall algebras, BPS web algebras and Harvey-Moore's algebra of BPS states.

\begin{figure}
\centering
\includegraphics[width=0.3\textwidth]{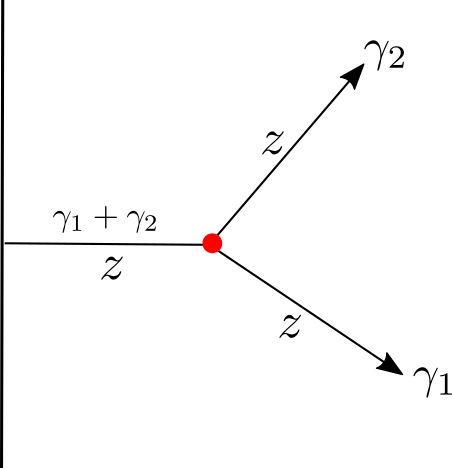}
\caption{A bulk vertex in the three-particle chamber of the $(A_1, A_2)$ theory leads to a web diagram that gives rise to the differential $Q(\alpha_{\gamma_1+\gamma_2}(z)) = \alpha_{\gamma_2}(z) \alpha_{\gamma_1}(z)$. }
\label{vertex}
\end{figure}

\paragraph{} The formulas in the three-particle chamber require us to include a trivalent ``interaction vertex'' between the three types of particles. Such a vertex is needed to give the differential \bea Q\big(\alpha_{\gamma_1+\gamma_2}(z) \big) = \alpha_{\gamma_2}(z) \alpha_{\gamma_1}(z) \eea a web-like interpretation. The differential arises from a process where the trivalent vertex associated to fan of BPS particles of charges $$\gamma_1, \gamma_2, -(\gamma_1+\gamma_2)$$ absorbs the BPS particle of charge $\gamma_1 +\gamma_2$ and emits two particles of charges $\gamma_2$ and $\gamma_1$, as depicted in the ``web" in Figure \ref{vertex}. We notice that the ghost number of a bulk local operator of the schematic form \bea \alpha_{\gamma_1} \, \alpha_{\gamma_2} \, \alpha_{-(\gamma_1+\gamma_2)}\eea is $+3$ whereas the spin is $+1$, which are precisely the required dimensions for a sensible interaction term in a holomorphic-topological theory in four dimensions\footnote{For a general theory with $d$ real topological directions and $k$ complex holomorphic directions we can deform the action by a local operator of ghost number $d+k$ and spin $k$.}.  

\begin{figure}[H]
\centering
\includegraphics[width=0.9\textwidth]{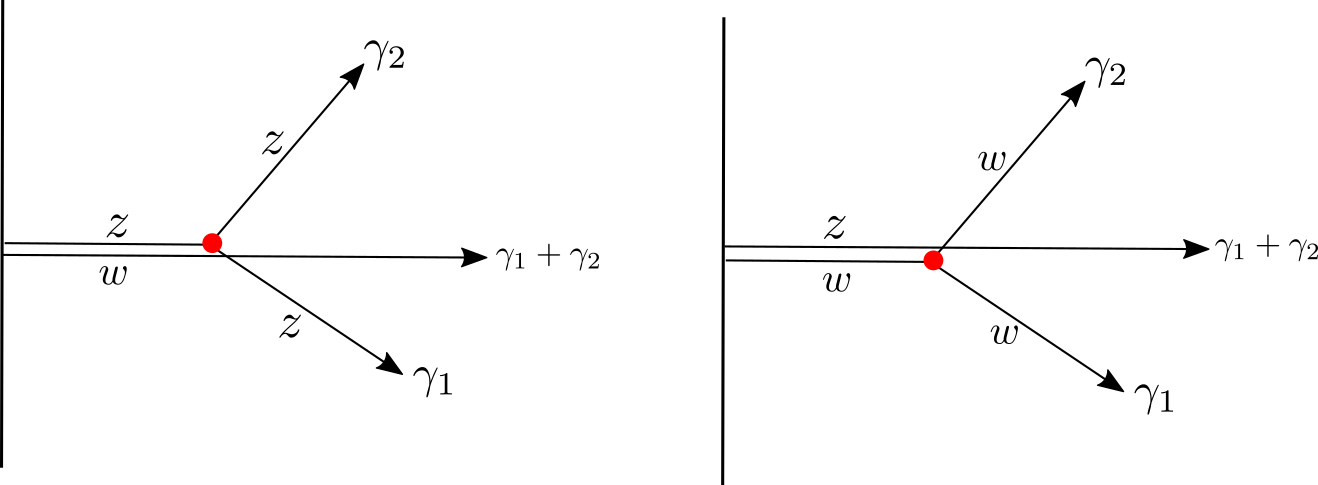}
\caption{The two web diagrams involving both crossings and bulk vertices giving rise to the two terms in the formula for $Q\big(\alpha_{\gamma_1+\gamma_2}(z) \alpha_{\gamma_1+\gamma_2}(w) \big)$.}
\label{vertex2}
\end{figure}

\paragraph{} The differential acting on a more general state, for instance \bea Q\big(\alpha_{\gamma_1+\gamma_2}(z) \alpha_{\gamma_1 + \gamma_2}(w) \big) = \frac{1}{z-w} \alpha_{\gamma_2}(z) \alpha_{\gamma_1+\gamma_2}(w) \alpha_{\gamma_1}(z) + \frac{1}{w-z} \alpha_{\gamma_2}(w) \alpha_{\gamma_1+\gamma_2}(z) \alpha_{\gamma_1}(w) \eea can possibly be interpreted as summing over diagrams where one $\gamma_1+ \gamma_2$-particle is absorbed by the bulk vertex with the other $(\gamma_1+\gamma_2)$-particles crossing through one of the edges of the bulk vertex giving us an R-matrix. For instance the above formula comes from a sum of two diagrams of this type as shown in Figure \ref{vertex2}. The more general formula simply is a sum of $n$ diagrams where the $i$th term corresponds to the absorption of the $i$th particle with the remaining $n-1$ particles pass through giving us a factor of $n-1$ R-matrices of the type \bea \prod_{j\neq i}\frac{1}{z_i-z_j}.\eea Fortunately this simple prescription leads to a nilpotent differential \eqref{qkoszul} that is free of singularities.

\paragraph{} We do not attempt here to give a full web interpretation for the product in the three-particle presentation of the algebra. 
The naive crossing terms would give a singular answer due to $(z-w)^{-1}$ factors, which is eliminated by extra terms whose interpretation presumably involve interaction vertices. 

\pagebreak

\appendix
\section{The Holomorphic-Topological Twist} \label{httwist}
\paragraph{} Here we refer as ``twisting'' of a SQFT the procedure of promoting a nilpotent supercharge $Q$ to a differential and passing to its cohomology. This operation produces a new, simpler theory. All translations which are $Q$-exact become trivial. 
Here we are interested in the Holomorphic-Topological twist of 4d ${\cal N}=2$ SQFTs \cite{Kapustin:2006hi}, leading to a twisted theory which is holomorphic in one complex direction $z$ and topological in the other two real directions $x^1$, $x^2$. Such a theory is only sensitive to a ``transverse holomorphic foliation'' of space-time, i.e. a projection of the 4d space-time to a complex one-dimensional holomorphic curve.  

\paragraph{} The twisted theory has a homological degree/ghost number which combines the ghost number and the Cartan generator of the $SU(2)_R$ R-symmetry of the physical theory. There is also a twisted rotation generator which combines 
the Cartan generators of rotations and $SU(2)_R$ R-symmetry of the physical theory and acts on the holomorphic direction of the twisted theory. 

\paragraph{} The HT twist of a 4d Lagrangian theory can be expressed conveniently in terms of super-fields which are forms in space-time, with differentials $\text{d} \bar z$ and $\text{d}x^i$ playing the role of super-space directions. In particular, it is easy to verify that the HT twist of a single hypermultiplet gives the HT version of a $\beta \gamma$ system, with action \bea \int_{\mathbb{R}^2 \times \mathbb{C}} \text{d}z \, \beta ( \text{d}+ \bar \partial) \gamma. \eea Here $\beta$ and $\gamma$ are bosonic superfields/forms of ghost number $1$ and twisted spin $\frac12$ \bea \beta = \sum_{i=0}^{3} \beta^{(i)},\,\,\,\,\,\,\, \gamma = \sum_{i=0}^{3} \gamma^{(i)}, \eea whose leading components $\beta^{(0)}, \gamma^{(0)}$ are the complex scalar fields in the hypermultiplet. The HT differential $\text{d}+ \bar \partial$ combines the de Rham differential in the topological plane and the Dolbeault differential in the holomorphic plane. 

\paragraph{} Bulk local operators are thus polynomials in holomorphic derivatives $\partial^n \beta$ and $\partial^n \gamma$. The index $I(\fq)$ counting local operators graded by their spin coincides with the Schur index
\begin{equation}
	I(\fq) = \frac{1}{\prod_{n=0}^\infty (1+x \fq^{2n+1})(1+ x^{-1} \fq^{2n+1})} = \Phi(x)^{-1}\Phi(x^{-1})^{-1} \, ,
\end{equation}
where $x$ is a flavour fugacity and we use $q = \fq^{2}$ as a spin fugacity. 

\paragraph{} The HT twist is also compatible with HT boundary conditions, which can arise as the twist of half-BPS boundary conditions for the physical theory. The simplest half-BPS boundary conditions for the hypermultiplet results in $\beta=0$ or $\gamma=0$ boundary conditions for the HT theory. Suppose we choose the latter. Then the index $II(\fq)$ counting boundary local operators graded by their spin coincides with the Schur half-index
\begin{equation}
	I\!\!I(\fq) = \frac{1}{\prod_{n=0}^\infty (1+ x \fq^{2n+1})} = \Phi(x)^{-1}\, .
\end{equation}
and counts polynomials in the bosonic generators $\partial^n \beta$.

\end{document}